\documentclass[aps,showpacs,twocolumn,superscriptaddress]{revtex4}

\usepackage{graphicx}
\usepackage{dcolumn}
\usepackage{bm}
\usepackage{amsmath}

\begin{document}

\preprint{}

\title{Detrended Cross-Correlation Analysis Consistently Extended to
Multifractality}

\author{Pawe\l{}~O\'swi\c ecimka}
\email{pawel.oswiecimka@ifj.edu.pl}
\affiliation{Institute of Nuclear Physics, Polish Academy of Sciences,
Krak\'ow, Poland.}

\author{Stanis\l{}aw~Dro\.zd\.z}
\affiliation{Institute of Nuclear Physics, Polish Academy of Sciences,
Krak\'ow, Poland.}
\affiliation{Faculty of Physics, Mathematics and Computer Science, Cracow
University of Technology, Krak\'ow, Poland.}

\author{Marcin~Forczek}
\author{Stanis\l{}aw Jadach}
\author{Jaros\l{}aw~Kwapie\'n}

\affiliation{Institute of Nuclear Physics, Polish Academy of Sciences,
Krak\'ow, Poland.}

\date{\today}

\begin{abstract}

We propose a novel algorithm - Multifractal Cross-Correlation Analysis (MFCCA)
- that constitutes a consistent extension of the Detrended Cross-Correlation
Analysis (DCCA) and is able to properly identify and quantify subtle
characteristics of multifractal cross-correlations between two time series.
Our motivation for introducing this algorithm is that the already existing
methods like MF-DXA have at best serious limitations for most of the signals
describing complex natural processes and often indicate multifractal
cross-correlations when there are none. The principal component of the present
extension is proper incorporation of the sign of fluctuations to their
generalized moments. Furthermore, we present a broad analysis of the model
fractal stochastic processes as well as of the real-world signals and show
that MFCCA is a robust and selective tool at the same time, and therefore
allows for a reliable quantification of the cross-correlative structure of
analyzed processes. In particular, it allows one to identify the boundaries of
the multifractal scaling and to analyze a relation between the generalized
Hurst exponent and the multifractal scaling parameter $\lambda_q$. This
relation provides information about character of potential multifractality in
cross-correlations and thus enables a deeper insight into dynamics of the
analyzed processes than allowed by any other related method available so far.
By using examples of time series from stock market, we show that financial
fluctuations typically cross-correlate multifractally  only for relatively
large fluctuations, whereas small fluctuations remain mutually independent
even at maximum of such cross-correlations. Finally, we indicate possible
utility of MFCCA to study effects of the time-lagged cross-correlations.

\end{abstract}


\maketitle

\section{Introduction}

Analysis of time series with nonlinear long-range correlations is often
grounded on a study of their multifractal
structure~\cite{halsey86,mandelbrot96,muzy94,kwapien12,struzik02,oswiecimka05,drozdz09,mandelbrot89}.
Existing algorithms used in such an analysis allow for determining generalized
fractal dimensions or H\"older exponents based either on statistical
properties of time series~\cite{kantelhardt02,grech13} or on time-frequency
information~\cite{muzy94,oswiecimka06}. Because of implementation simplicity
and their utility, these algorithms have already been applied to characterize
correlation structure of data in various areas of science like
physics~\cite{muzy08,subramaniam08},
biology~\cite{ivanov99,makowiec09,rosas02},
chemistry~\cite{stanley88,udovichenko02}, geophysics~\cite{witt13,telesca05},
economics~\cite{ausloos02,calvet02,turiel05,drozdz10,oswiecimka08,zhou09,bogachev09,su09},
hydrology~\cite{koscielny06}, atmospheric physics~\cite{kantelhardt06},
quantitative linguistics~\cite{ausloos12,grabska12},
music~\cite{jafari07,oswiecimka11}, and human communications~\cite{perello08}. As an important step towards quantifying
complexity, in recent years algorithms designed for investigation of fractal
cross-correlations were proposed~\cite{podobnik08,podobnik09} followed by the
new statistical cross-correlation tests~\cite{zebende11,podobnik11}. These
developments are based on the Detrended Cross-Correlation Analysis (DCCA)
which constitutes a straightforward generalization of the fractal
auto-correlation (DFA)~\cite{kantelhardt01} on the case of fractally
cross-correlated signals. In that case, the cross-correlation scaling exponent
$\lambda$ can be obtained. However, literature still lacks comprehensive
interpretation of this quantity.

Subsequently, the multifractal extension (MF-DXA) of the DCCA method was
proposed~\cite{zhou08}. Other closely related methods to deal with
multifractal cross-correlations have also been introduced~\cite{kristoufek11}.
However, these extensions naturally involve computation of arbitrary powers of
cross-covariances and this leads to serious limitations since such
cross-covariances may, in general, become negative. In such a case the net
result, expressed in terms of the usual fluctuation functions, thus becomes
complex-valued which does not allow to determine the scaling exponents by
conventional means. A simplistic resolution, so far available in the
literature, to this difficulty is based on taking
modulus~\cite{jiang11,he11,li12,wang12} of the cross-covariance function in
order to get rid of its negative signs. In most realistic cases, as our
analysis below shows, this however seriously distorts or even spuriously
amplifies the multifractal cross-correlation measures. Our motivation
therefore is to elaborate an algorithm that we call Multifractal
Cross-Correlation Analysis (MFCCA), such that for any two signals it allows to
compute their arbitrary-order covariance function and at the same time it
properly takes care of the relative signs in the signals.

The proposed method allows us to calculate the spectrum of the exponents
$\lambda_q$, which characterize multifractal properties of the
cross-covariance. However, unlike the method proposed ealier, in our
procedure, the scaling properties of the $q$th order cross-covariance function
are estimated with respect to the original sign of the cross-covariance. This
procedure makes the method both more sensitive to cross-correlation structure
and free from limitations of other algorithms. It also turns out that the
proposed method is a more natural generalization of the monofractal DCCA than
is MF-DXA. The robustness of our algorithm makes it applicable to different
data types in various fields of science.

\section{Description of the MFCCA algorithm \label{Method}}

Multifractal Cross-Correlation Analysis consists of several steps which are
described in detail below. As it was mentioned above, MFCCA has been developed
based on the DCCA procedure \cite{podobnik08}, therefore the initial steps are
the same.

Consider two time series $x_i$, $y_i$ where $i=1,2...N$. At first, the signal
profile has to be calculated for each of them:
\begin{equation}
X\left(j\right) =\sum_{i=1}^j[x_{i}-\langle x\rangle] ,\quad
Y\left(j\right) =\sum_{i=1}^j[y_{i}-\langle y\rangle].
\end{equation}
Here, $\langle \rangle$ denotes averaging over entire time series. Then, both
signal profiles are divided into $M_s=N/s$ disjoint segments $\nu$ of length
$s$. For each box $\nu$, the assumed trend is estimated by fitting a
polynomial of order $m$ ($P^{(m)}_{X,\nu}$ for $X$ and $P^{(m)}_{Y,\nu}$ for
$Y$). Based on our own experience~\cite{oswiecimka06}, as optimal we use a
polynomial of order $m=2$ throughout this paper but the proposed procedure is
not restricted to this particular order and can be used for much larger one
when needed (as, for instance, in signals involving a highly periodic
component~\cite{horvatic11, ludescher11}). Next, the trend is subtracted from
the data and the detrended cross-covariance within each box is calculated:
\begin{multline}
F_{xy}^{2}(\nu,s)=\frac{1}{s}\Sigma_{k=1}^{s}\lbrace
(X((\nu-1)s+k)-P^{(m)}_{X,\nu}(k)) \times \\ \times
(Y((\nu-1)s+k)-P^{(m)}_{Y,\nu}(k))\rbrace
\label{Fxy2}
\end{multline}
In contrast to the detrended variance calculated in the MFDFA
procedure~\cite{kantelhardt02}, in the present case, $F_{xy}^{2}(\nu,s)$ can
take both positive and negative values (for an example see Sec.~\ref{stocks}
Fig.\ref{fig12}). Therefore, gradual investigation of scaling properties from
small to large fluctuations through their covariances of increasing order
should take into account also sign of $F_{xy}^{2}(\nu,s)$. Accordingly, the
most natural form of the $q$th order covariance function is postulated by the
following equation:
\begin{equation}
F_{xy}^{q}(s)=\frac{1}{M_s}\Sigma_{\nu=1}^{M_s} {\rm
sign}(F_{xy}^{2}(\nu,s))|F_{xy}^{2}(\nu,s)|^{q/2} ,
\label{Fq}
\end{equation}
where ${\rm sign}(F_{xy}^{2}(\nu,s))$ denotes the sign of $F_{xy}^{2}(\nu,s)$.
The parameter $q$ can take any real number except zero. However, for $q=0$,
the logarithmic version of Eq.~(\ref{Fq}) can be employed~\cite{kantelhardt02}:
\begin{equation}
F_{xy}^{0}(s)=\frac{1}{M_s}\Sigma_{\nu=1}^{M_s} {\rm sign}(F_{xy}^{2}(\nu,s))
\ln|F_{xy}^{2}(\nu,s)| .
\end{equation}
As we can see in Eq.~(\ref{Fq}), for negative values of $q$, small values of
the covariance function $F_{xy}^{2}(\nu,s)$ are amplified, while for large
$q>0$, its large values dominate. Moreover, the formula for calculating
$F_{xy}^{q}(s)$ respects the genuine signs of the amplified (or supressed)
fluctuations of the detrended cross-covariance function (Eq.~(\ref{Fxy2}))
and, at the same time, it allows to avoid complex numbers associated with the
arbitrary powers of negative fluctuations. The above described steps of MFCCA
should be repeated for different scales $s$. If the so-obtained function
$F_{xy}^{q}(s)$ does not develop scaling, by for instance fluctuating around
zero, there is no fractal cross-correlation between the time series under
study for the considered value of $q$. Multifractal cross-correlation is
expected to manifest itself in the power-law dependence of $F_{xy}^{q}(s)$ (if
the $q$th order covariance function is negative for every $s$, we may take
$F_{xy}^{q}(s)\longrightarrow -F_{xy}^{q}(s)$~\cite{podobnik08}) and the
following relation is fulfilled:
\begin{equation}
F_{xy}^{q}(s)^{1/q}=F_{xy}(q,s) \sim s^{\lambda _q}
\label{Fxy}
\end{equation}
(or $\exp(F_{xy}^{0}(s)) = F_{xy}(0,s) \sim s^{\lambda_0} $ for $q=0$), where
$\lambda_q$ is an exponent that quantitatively characterizes fractal
properties of the cross-covariance. For the monofractal cross-correlation, the
exponents $\lambda _q$ are independent of $q$ and equal to $\lambda$ as
obtained from the DCCA method. In the case of multifractal cross-correlation,
however, $\lambda _q$ varies with $q$, with $\lambda$ retrieved for $q=2$. The
minimum and maximum scales ($s_{min}$ and $s_{max}$, respectively) depend on
the length $N$ of the time series under study. In practice, it is reasonable
to take $s_{max} < N/5$.

\section{Analysis of examplary models and stock market data}

In order to verify the usefulness of MFCCA algorithm, we test it by using both
artificially generated cross-correlated time series and real-world signals. In
order to avoid divergent moments due to fat tails in the distribution of
fluctuations, we restrict $q$ to $\langle -4,4 \rangle$ with a step $0.2$
throughout this paper. In the case of computer-generated signals, results for
each process are averaged over its 20 independent realizations.

\subsection{ARFIMA processes}

\begin{figure}
\includegraphics[width=0.45\textwidth, height=0.4 \textwidth]{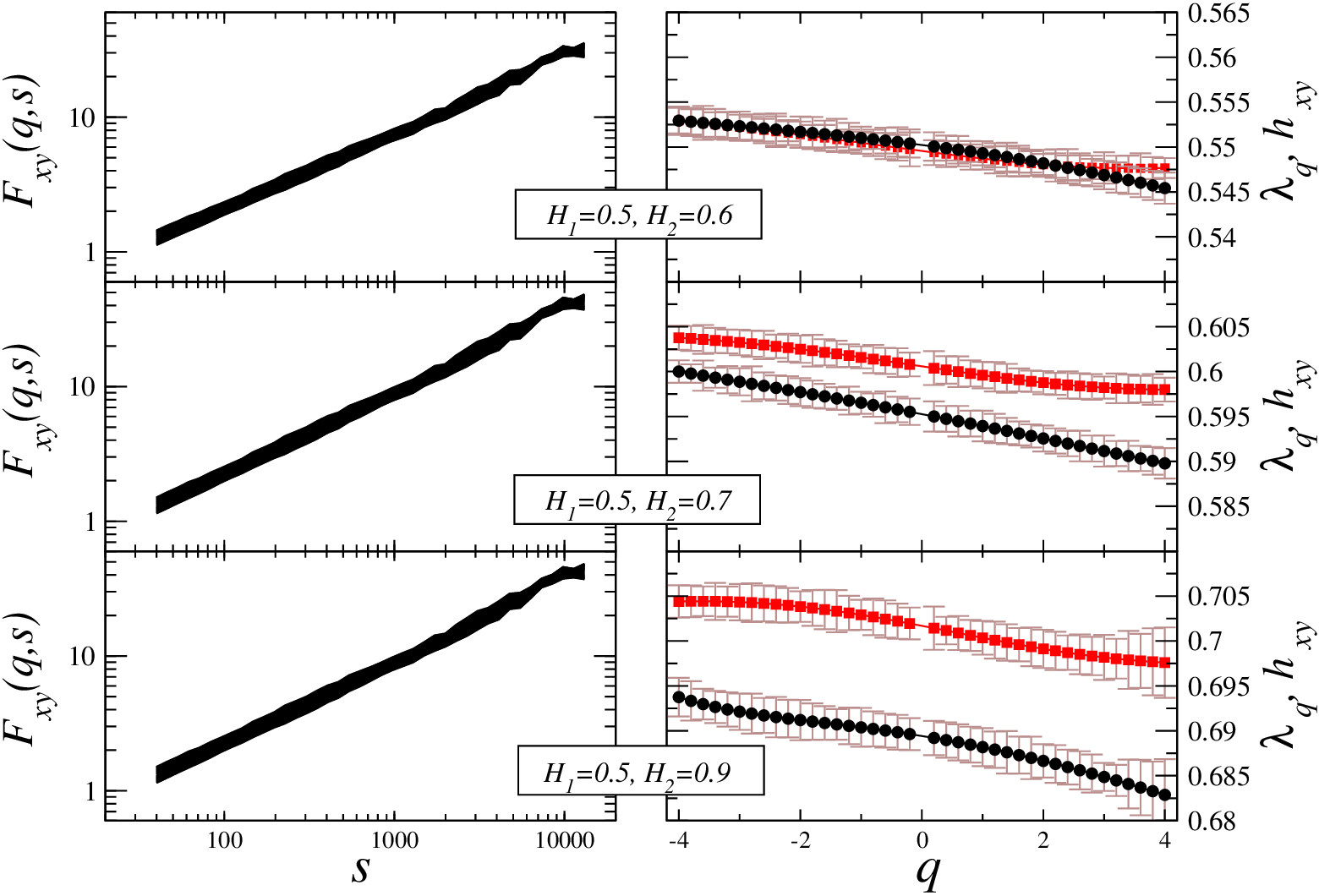}
\caption{(Color online) (Left) Family of the $q$th-order cross-covariance functions
$F_{xy}(q,s)$ calculated for ARFIMA processes for three different combinations
of the parameters $H_1$ and $H_2$. The lowest and the highest line in each
panel refers to $q=-4$ and $q=4$, respectively. (Right) Multifractal
cross-correlation scaling exponents $\lambda_q$ (black circles) and the
average generalized Hurst exponents $h_{xy}(q)$ (red squares). Error bars
indicate standard deviation calculated from 20 independent realizations of the
corresponding processes.}
\label{fig1}
\end{figure}

We start our study from an analysis of the well-known ARFIMA
processes~\cite{hosking81}, which are examples of monofractal, long-range
correlated signals. In Ref.~\cite{podobnik08}, such processes were used to
show usefulness of the DCCA algorithm. Our goal is to show the
cross-correlation structure of the above-mentioned processes more completely.
To generate a pair ($x_i,y_i$) of the cross-correlated ARFIMA processes, we
use the following equations:
\begin{equation}
x_i=\Sigma_{j=1}^{\infty}a_j(d_x)x_{i-j}+\epsilon_i \,
\label{x}
\end{equation}
\begin{equation}
y_i=\Sigma_{j=1}^{\infty}a_j(d_y)y_{i-j}+\epsilon_i
\label{y}
\end{equation}
where $d_x$ and $d_y$ are parameters characterizing linear long-range
autocorrelations of the times series. These quantities can be related to the
Hurst exponents~\cite{kantelhardt01} by the relation $H=1/2+d_{x(y)}$,
($-1/2<d_{x(y)}<1/2$). Positively correlated (persistent) time series are
characterized by $H>0.5$, whereas negative autocorrelation (antipersistent
signal) is characterized by $H<0.5$; H=0.5 means no linear autocorrelation.
The quantity $a_j(d_{x(y)})$ is called weight and is defined by $a_j(d_{x(y)})
= \Gamma (j-d_{x(y)})/[\Gamma(-d_{x(y)})\Gamma(1+j)]$, where $\Gamma()$ stands
for Gamma function. $\epsilon_i$ is an i.i.d.~Gaussian random variable. The
processes $x_i$ and $y_i$ are cross-correlated, because the same noise
component $\epsilon_i$ is used in both Eq.(\ref{x}) and Eq.(\ref{y}). We
generate three pairs of cross-correlated signals: ($H_1$=0.5, $H_2$=0.6),
($H_1$=0.5, $H_2$=0.7), and ($H_1$=0.5, $H_2$=0.9), where $H_1$ and $H_2$
characterize long-range autocorrelation of the first and the second time
series, respectively. In order to obtain statistically significant results, we
generate time series of lengh $N=100,000$ points each.

In the left panels of Fig.~\ref{fig1}, we present the calculated $F_{xy}(q,s)$
for all the signal pairs. Each line corresponds to a different value of $q$.
As it can be seen, in all the cases, $F_{xy}(q,s)$ is a power function of
scale $s$. This indicates the fractal nature of the cross-correlations.
Moreover, for all types of signals, the functions $F_{xy}(q,s)$ are almost
parallel to each other implying largely homogeneous character of the
corresponding cross-correlations. Indeed, as shown in the right panels of the
Fig.~\ref{fig1}, the difference between the extreme values of $\lambda_q$
expressed by $\Delta \lambda_q=\max(\lambda_q)-\min(\lambda_q)$ is
approximately 0.005, 0.007, and 0.011 for the top, middle, and the bottom
panel, respectively. These narrow ranges of $\lambda_q$ indicate that the
ARFIMA processes reveal correlations that are monofractal regardless of the
types of linear autocorrelation of signals.

In literature, the estimated fractal cross-correlations are often related to
the fractal properties of the individual signals
themselves~\cite{podobnik08,siqueira10,shadkhoo09}. Therefore, in
Fig.~\ref{fig1}, we also show the average of the generalized Hurst exponents
~\cite{kantelhardt02}:
\begin{equation}
h_{xy}(q)=(h_x(q)+h_y(q))/2,
\end{equation}
where $h_x(q)$ and $h_y(q)$ refer to fractal properies of individual time
series, respectively and, for $q=2$, they correspond to the Hurst exponent
$H$. It is worth noticing that relation between $\lambda_q$ and $h_{xy}(q)$
depends on temporal organization of the signals as determined by their Hurst
exponents. For two signals whose Hurst exponents $H$ are alike, their
multifractal cross-correlation characteristics described by $\lambda_q$ and
$h_{xy}(q)$ are almost identical, while the divergence between $\lambda_q$ and
$h_{xy}(q)$ becames more sizeble for time series with more significant
differences in autocorrelation (different Hurst exponent $H$).

This result means that, in the case of the ARFIMA processes, the relation
$\lambda\approx (h_x(2)+h_y(2))/2$ introduced in Ref.~\cite{podobnik08}
applies only to a situation when differences between $h_x$ and $h_y$ are
negligible.

\subsection{Markov-switching multifractal model \label{Markov}}

As an example of multfractal process, we consider the Markov-switching
multifractal model (MSM)~\cite{liu07}. MSM is an iterative model, which is
able to replicate hierarchical, multiplicative structure of real data and,
thus, insures multifractal properties of the generated time series. Because of
its properties, MSM is commonly used in finance, where multifractality of
price fluctuations is one of the main stylized facts~\cite{liu08,kwapien05}.
Equally well this model can be used to simulate many other multifractal time
series representing natural phenomena as it is able to generate the volatility
clustering responsible for the underlying nonlinear temporal
correlations~\cite{drozdz09}. Below, we present the main stages of the model's
construction.

\begin{figure}
\includegraphics[width=0.5 \textwidth, height=0.38 \textwidth, clip=true
]{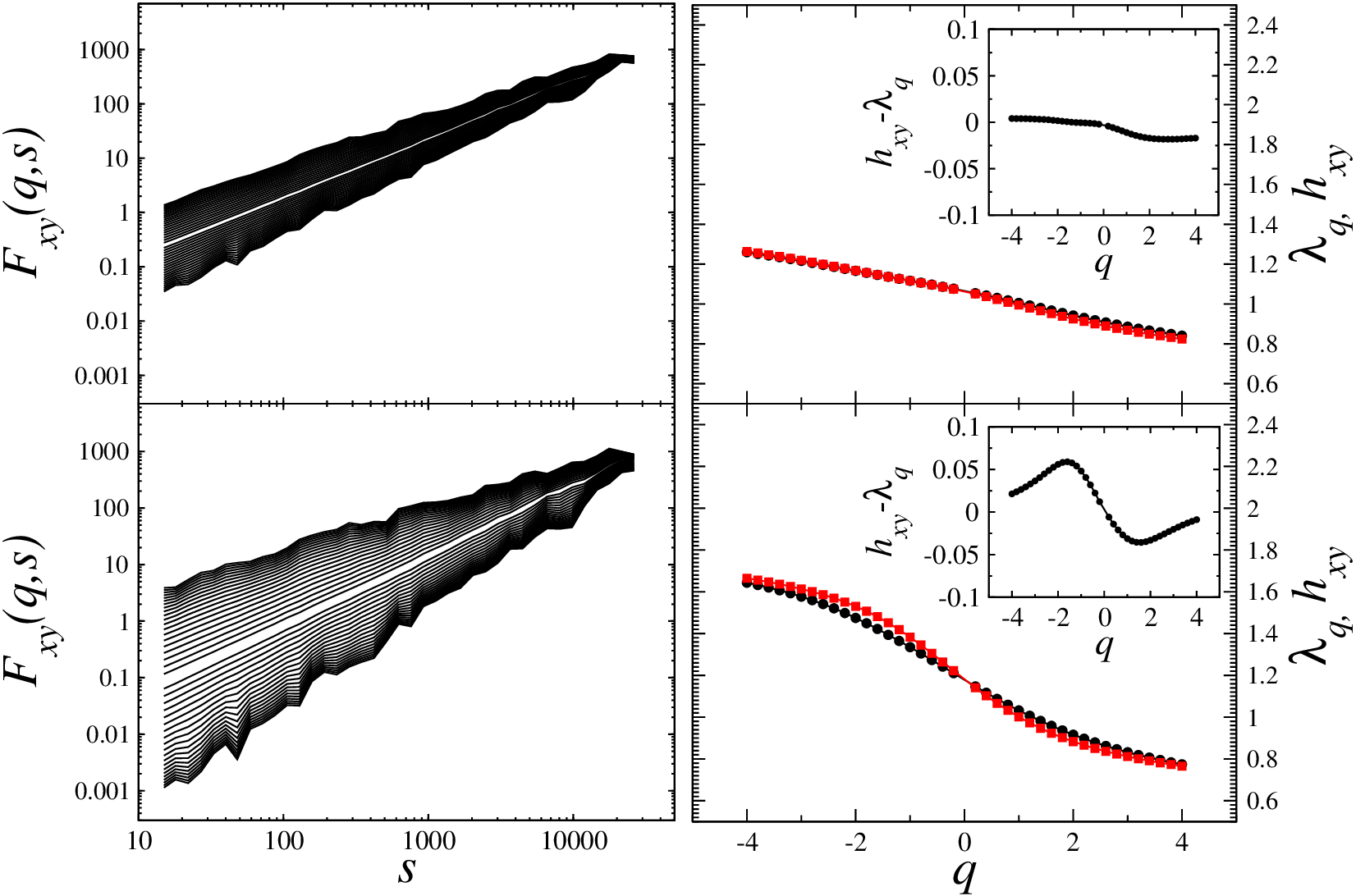}
\caption{(Color online) (Left) Family of the $q$th-order cross-covariance functions
$F_{xy}(q,s)$ calculated for two pairs of the MSM time series corresponding to
$m_{0}^{(1)}=1.2$, $m_{0}^{(2)}=1.35$ (top) and $m_{0}^{(1)}=1.2$,
$m_{0}^{(2)}=1.6$ (bottom). The lowest and the highest line on each panel
refers to $q=-4$ and $q=4$, respectively. (Right) Multifractal
cross-correlation scaling exponents $\lambda_q$ (black circles) and the
average generalized Hurst exponents $h_{xy}(q)$ (red squares). Insets present
the differences between $\lambda_q$ and $h_{xy}$.}
\label{fig2}
\end{figure}

\begin{figure}
\includegraphics[width=0.4\textwidth,height=0.37 \textwidth, clip=true]{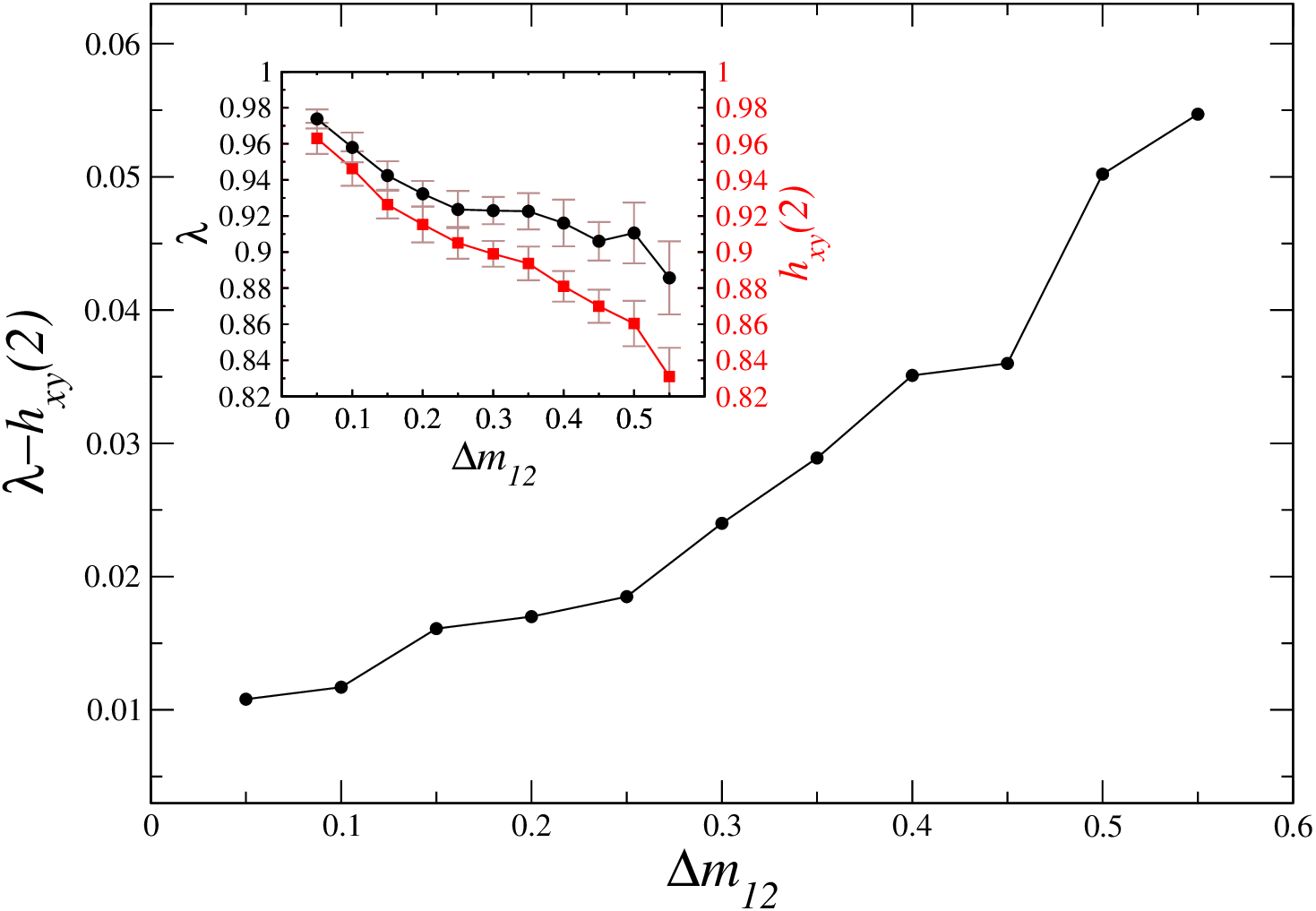}
\caption{(Color online) (Inset) Average $\lambda_2=\lambda$ and average $h_{xy}(2)$ as a
function of $\Delta m_{12}=m_{0}^{(2)}-m_{0}^{(1)}$. Black circles and red
squares correspond to the $\lambda$ and $h_{xy}(2)$, respectively. Error bars
indicate standard deviation calculated from 20 independent realizations of the
corresponding process. (Main) Difference of the average $\lambda$ and the
average $h_{xy}(2)$ as a function of $\Delta m_{12}$.}
\label{fig3}
\end{figure}

In MSM, evolution of an observable $r_t$ in time $t$ is modeled by the
formula~\cite{liu08}:
\begin{equation}
r_t=\sigma_t\cdot u_t ,
\label{msm}
\end{equation}
where $u_t$ stands for a Gaussian random variable and $\sigma_t$ (multifractal
process) stands for the instantaneous volatility component. The volatility
$\sigma_t$ is a product of $k$ multipliers $M_1(t),M_2(t),...,M_k(t)$ such that
\begin{equation}
\sigma^2_t=\sigma^2\prod _{i=1}^k M_i(t) ,
\end{equation}
where $\sigma^2$ is a constant factor. A common version of the model assumes
that the multipliers $M_i(t)$ are drawn from the binomial or from the
log-normal distribution. Here, we use the binomial one with $M_i(t) \sim
\{m_0,2-m_0\}$, $1 \leq m_0<2$. Any change of a multiplier in the hierarchical
structure of volatility is determined by the transition
probabilities~\cite{liu07}:
\begin{equation}
\gamma _i = 1-(1-\gamma _k)^{b^{i-k}}      ,\quad i=1,2...k.
\end{equation}
Thus, a multiplier $M_i(t)$ is renewed with probability $\gamma_i$ and remains
unchanged with probability $1-\gamma_i$. The parameter $\gamma_k$ is taken
from the range $(0,1)$, and $b > 1$. We put $\gamma_k=0.5$ and $b=2$, which
leads to the relation:
\begin{equation}
\gamma _i = 1-(0.5)^{2^{i-k}}      ,\quad i=1,2...k.
\end{equation}
Thus, for the initial stages of the cascade, a renewal of the multipliers
$M_i(t)$ occurs with relatively small probability, while the largest
$\gamma_i=0.5$ appears for $i=k$.

\subsubsection{Unsigned version of the MSM model}

For the purpose of this analysis, we generate a set of multifractal time
series ($\sigma_t$) of length $131, 072$ points each. However, in all
realizations of the model, we conserve the hierarchical structure of the
multipliers, since the renewals of $M_i(t)$ appear for the same $i$ and $t$ in
each generated series. This procedure insures cross-correlations between
series with different $m_0$.

In Fig.~\ref{fig2}, we present sample results of MFCCA obtained for two pairs
of MSM series with the parameters $m_{0}^{(1)}=1.2$ and $m_{0}^{(2)}=1.35$ (top
panels) and $m_{0}^{(1)}=1.2$ and $m_{0}^{(2)}=1.6$ (bottom panels). The
$q$th-order covariance functions $F_{xy}(q,s)$ (left hand side of this Figure)
display a clear multifractal scaling within the whole range $(-4,4)$ of the
$q$ values. The resulting $\lambda_q$ is a decreasing function of $q$, which
is a hallmark of multifractality. Moreover, the rate of decrease of
$\lambda_q$ depends on the values of mutlipliers. For the first pair of
signals (with $m_{0}^{(2)}=1.35$), the  exponents $\lambda_q$ are contained in
the range (0.81,1.25), while for the second pair (with $m_{0}^{(2)}=1.6$),
$0.75 \le \lambda_q \le 1.7$. In the same Fig.~\ref{fig2}, we also show the
average of the generalized Hurst exponents calculated for each time series
independently (red squares).

Interestingly, for the signals with a relatively small difference $\Delta
m_{12}=m_{0}^{(2)}-m_{0}^{(1)}$ - in other words, for similar multifractals -
$\lambda_q$ approximately equals the average of $h_{x}(q)$ and $h_y(q)$. A
tiny difference between $h_{xy}(q)$ and $\lambda_q$ is here visible only for
$q>0$. This effect is depicted more quantitatively in the insets of
Fig.~\ref{fig2}, where $h_{xy}(q)-\lambda_q$ is presented as a function of
$q$. The maximum deviation from zero can be seen for $q=2.2$, reaching a value
of $0.02$. For the second pair of signals, the difference
$h_{xy}(q)-\lambda_q$ is more pronounced and concerns both negative and
positive $q$'s. In this case, the largest difference of $h_{xy}(q)$ and
$\lambda_q$ is for $q=-2$ and equals 0.07.

\subsubsection{Relation between $\lambda_q$ and $h_{xy}$}

To have some insight into the relation between  $\lambda_q$ and $h_{xy}(q)$,
we perform a systematic MFCCA study for the set of time series pairs, such
that one of them is generated with $m_{0}^{(1)}=1.2$ and the other one with
$m_{0}^{(2)}$ from the range $\langle 1.25, 1.9 \rangle$ (the step is 0.05).
However, the multifractal characteristics were possible to estimate only for
$\Delta m_{12}<0.6$. In the case of $\Delta m_{12}>0.6$, $F_{xy}^q$ takes both
positive and negative values and Eq.(\ref{Fxy}) is not satisfied. At first, we
focus on the relationship between $\lambda_2=\lambda$ and the average Hurst
exponent $h_{xy}(2)$. In the inset of Fig.~\ref{fig3}, we present these
quantities as a function of $\Delta m_{12}$. It is clearly visible that both
these quantities are monotonically decreasing and they take approximately the
same values for small $\Delta m_{12}$. However, for $\Delta m_{12}>0.25$,
$\lambda(\Delta m_{12})$ decreases slower than the Hurst index (thus
$\lambda>H$) and the statistics diverge. To highlight this result, we
calculate also the difference between these two quantities which is shown in
Fig.\ref{fig3}. As one can see, $\lambda-h_{xy}(2)$ is an increasing function
of $\Delta m_{12}$. This result indicates that the difference between
$\lambda$ and the average Hurst exponent becomes larger for time series whose
multifractal characteristics depart more from each other, while the opposite
is observed when these characteristics are alike, which at the same time
results in stronger cross-correlations.

\begin{figure}
\includegraphics[width=0.46 \textwidth,height=0.41 \textwidth]{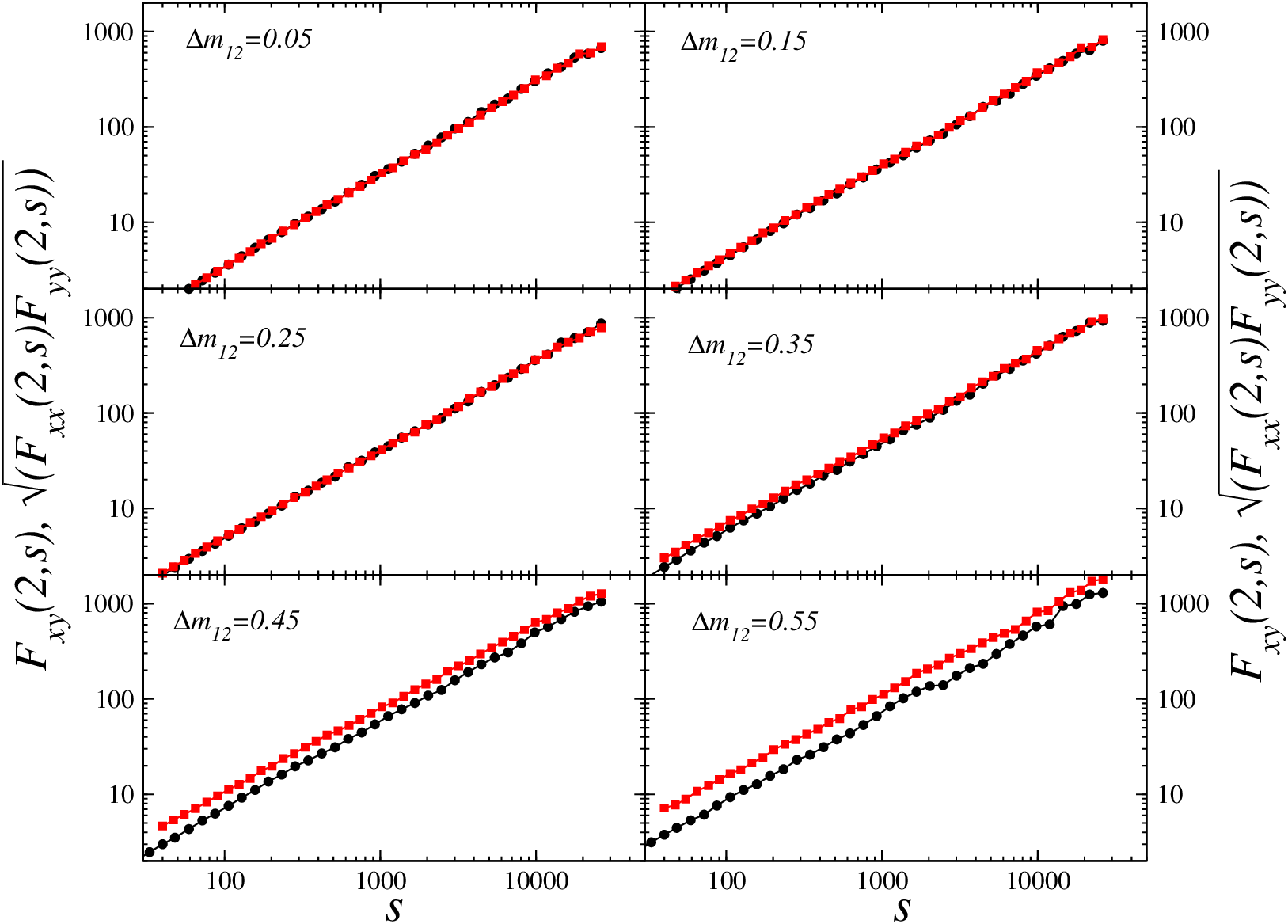}
\caption{(Color online) Comparision of covariance function $F_{xy}(2,s)$ (black circles) with
its equivalent $\sqrt{F_{xx}(2,s)F_{yy}(2,s)}$ (red squares) derived from the
variance functions of individual MSM time series. The slope of this functions
refers to $\lambda_2$ and $h_{xy}(2)$, respectively.}
\label{fig4}
\end{figure}

\begin{figure}
\includegraphics[width=0.44 \textwidth,height=0.35 \textwidth, clip=true]{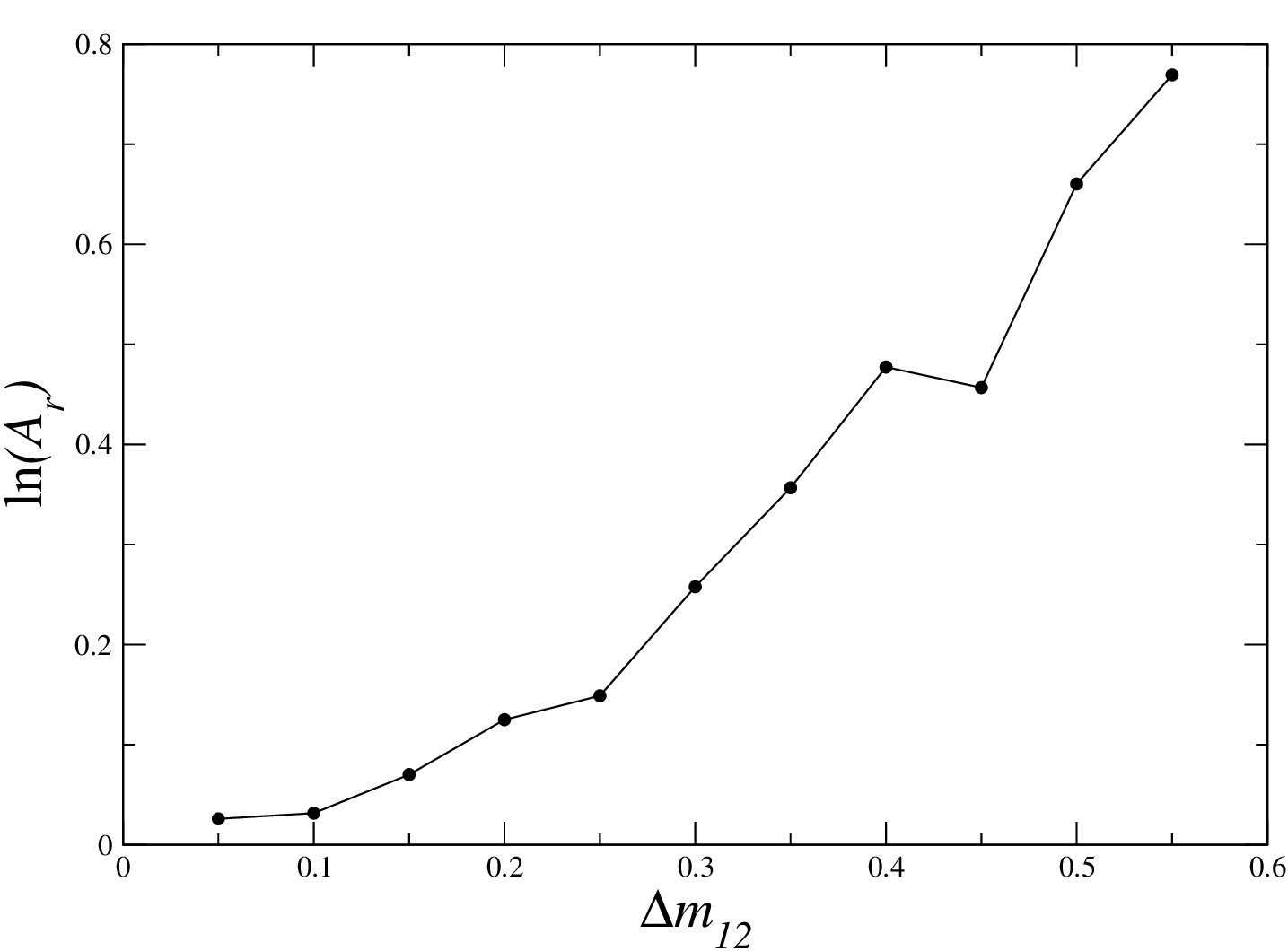}
\caption{Plot of $\ln(A_r)$ calculated for $q=2$ as a function of similarity
parameter $\Delta m_{12}$ of the MSM time series.}
\label{fig5}
\end{figure}

\begin{figure}
\includegraphics[width=0.45\textwidth,height=0.4 \textwidth, clip=true]{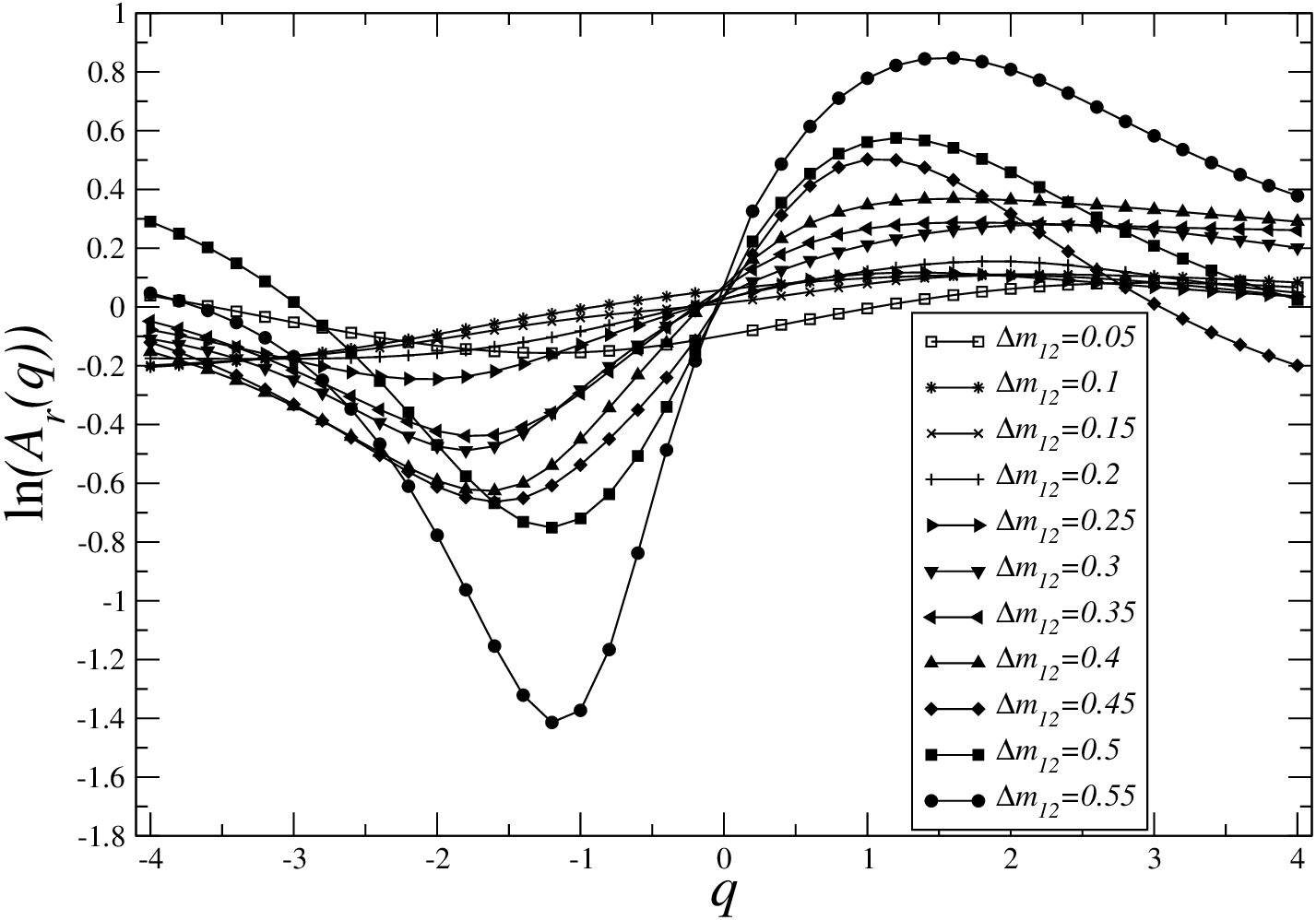}
\caption{Plot of $\ln(A_r)$ as a function of $q$. Each line corresponds to
different degree of similarity $\Delta m_{12}$ of the MSM time series.}
\label{fig6}
\end{figure}

To better understand this effect, we analyze a covariance function
$F_{xy}(2,s) \sim s^\lambda$ and an expression based on fluctuation
functions~\cite{kantelhardt02}: $\sqrt{F_{xx}(2,s)F_{yy}(2,s)} \sim
s^{\frac{h_x(2)+h_y(2)}{2}} = s^{h_{xy}(2)}$. In Fig.~\ref{fig4}, we show
these functions calculated for different values of $\Delta m_{12}$. It is easy
to notice that the presented functions are almost identical to each other for
small $\Delta m_{12}$. However, the larger $\Delta m_{12}$ is, the more
visible is a departure between the analyzed statistics. In all cases, the
values of $F_{xy}(2,s)$ are at most equal to $\sqrt{F_{xx}(2,s)F_{yy}(2,s)}$
and estimated $\lambda$ is larger than $h_{xy}(2)$. These numerical results
are in accord with the following relation:
\begin{equation}
F_{xy} (2,s)\leq \sqrt{F_{xx}(2,s)F_{yy}(2,s)} ,
\label{FxyF}
\end{equation}
which straightforwardly results from the definitions of these quantities
considered in terms of the scalar products of vectors formed from the
underlying time series~\cite{he11}. In order to more clearly see the
relationship between $\lambda$ and $h_{xy}(2)$, we can reformulate
Eq.~(\ref{FxyF}) in the case when the relations $F_{xy}
(2,s)=a_{xy}s^\lambda$, $F_{xx}(2,s)=a_xs^{h_x(2)}$, and
$F_{yy}(2,s)=a_ys^{h_y(2)}$ apply, to obtain:
\begin{equation}
a_{xy}s^{\lambda}\leq (a_xa_y)^{1/2}s^{\frac{h_x(2)+h_y(2)}{2}} .
\end{equation}
This leads to:
\begin{equation}
\lambda \leq \log_s(\frac{(a_xa_y)^{1/2}}{a_{xy}}) + \frac{h_x(2)+h_y(2)}{2}.
\end{equation}
For two identical time series, the equality in Eq.~(\ref{FxyF}) holds leading
to obvious $\lambda = \frac{h_x(2)+h_y(2)}{2}$. In general, however,
\begin{equation}
A_r = \frac{(a_xa_y)^{1/2}}{a_{xy}}\neq 1,
\end{equation}
and thus a difference between $\lambda$ and $h_{xy}(2)$ in either direction is
allowed or even forced, depending on a sign of $\log_s(A_r)$. For negative
values of this quantity, $\lambda$ has to be smaller than $h_{xy}(2)$, while
for positive values it can become larger. An example demonstrating the rate of
changes of $\ln(A_r)$ as a function of $\Delta m_{12}$ for $q=2$ is shown in
Fig.~\ref{fig5}. In this case, $\ln(A_r)$ is positive and quickly increases
with $\Delta m_{12}$, thus with the degree of dissimilarity between the two
series. The related dependencies are even more involved and appear to strongly
vary with the parameter $q$ as it is more systematically shown in
Fig.~\ref{fig6}. The $\ln(A_r)$ is seen to be positive for $q > 0$ with an
increasing value at maximum with increasing $\Delta m_{12}$, and a larger
amplitude of changes with increasing $q$. Similar, but reversed in sign and
with an even larger amplitude of changes, is the situation for $q < 0$. These
results nicely coincide - and thus point to their origin - with those
presented in Fig.~\ref{fig2}, where $\lambda_q$ is larger than $h_{xy}(q)$ for
positive values of $q$ and smaller for negative ones. Even the maxima of these
differences occur for those values of $q$, where they are seen in
Fig.~\ref{fig6} and they are larger on the negative side of $q$. Of course,
they are also larger for larger $\Delta m_{12}$.

\begin{figure}
\includegraphics[width=0.45\textwidth,height=0.4 \textwidth, clip=true]{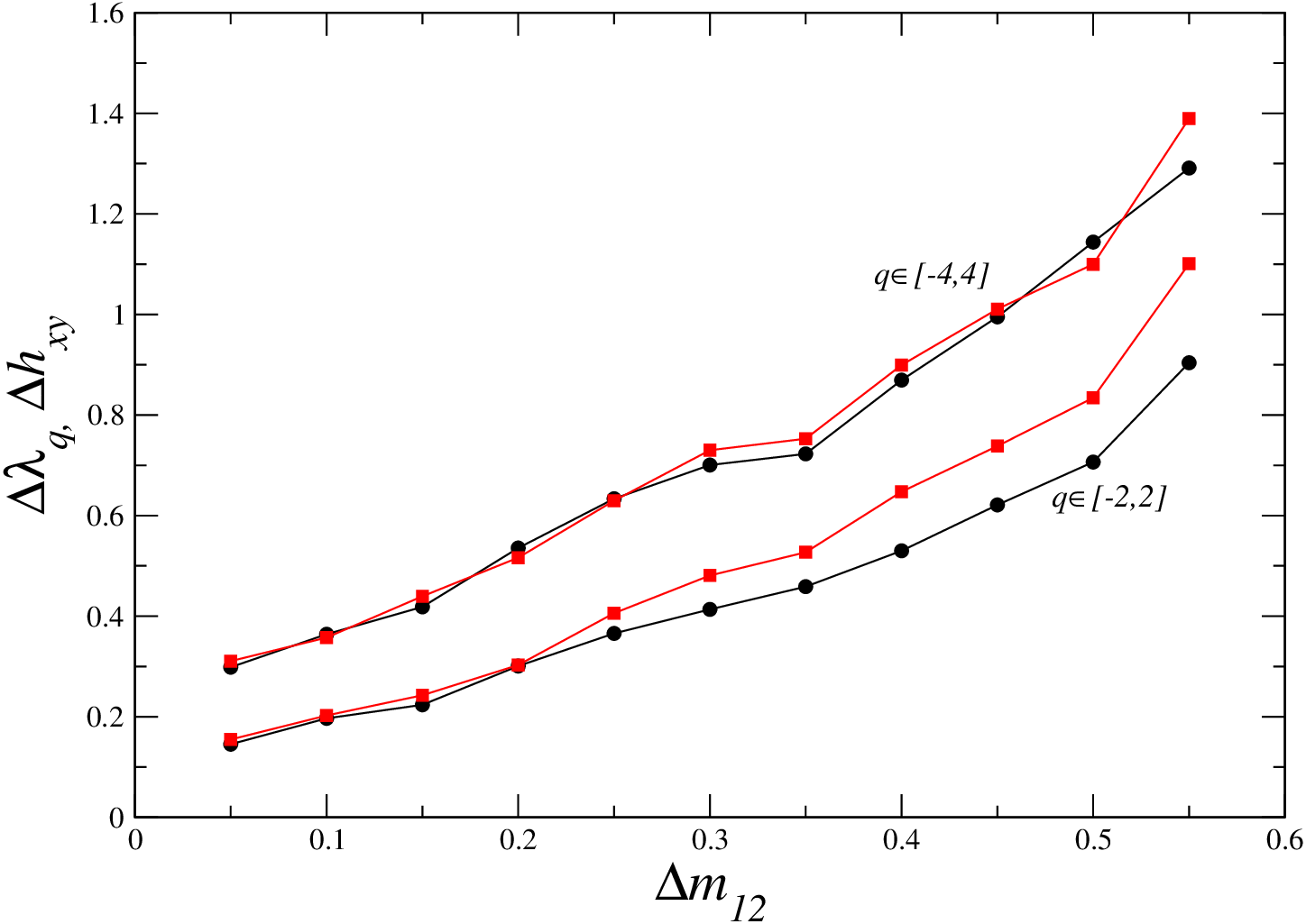}
\caption{(Color online) $\Delta \lambda_q$ (black circles) and $\Delta h_{xy}$ (red squares)
as functions of multiplier difference $\Delta m_{12}=m_{0}^{(2)}-m_{0}^{(1)}$
calculated for two ranges of the parameter $q$.}
\label{fig7}
\end{figure}

The difference between $\lambda_q$ and $h_{xy}(q)$ has its reflection - also
consistent with the findings presented in Figs.~\ref{fig2} and~\ref{fig6} - in
another popular multifractal measure, namely in the range of scaling
exponents. In Fig.~\ref{fig7}, we display $\Delta \lambda_q$ as a function of
$\Delta m_{12}$ for the two ranges of $q$: $-2 \le q \le 2$ and $-4 \le q \le
4$. For comparison, in the same Figure, we show $\Delta
h_{xy}=\max(h_{xy})-\min( h_{xy})$ calculated for the same ranges of $q$.  The
$\Delta h_{xy}(q)$ and $\Delta \lambda_q$ are seen to be monotonically
increasing functions of $\Delta m_{12}$ in all the cases. However, for $-4 \le
q \le 4$ these characteristics are almost the same, while for $-2 \le q \le 2$
the difference between $\Delta h_{xy}$ and $\Delta \lambda_q$ systematically
increases with $\Delta m_{12}$. This suggests that for relatively large values
of $|q|$ (magnifying the largest and the smallest fluctuations of
instantaneous volatility components) the fractal character of the considered
processes is similar, which may reflect the effect of preserving the same
hierarchical structure of multipliers for all generated multifractals, where
only relative changes of volatility are possible. The above results thus
indicate that the difference between $h_{xy}(q)$ and $\lambda_q$ is to be
considered an important ingredient of measure of the fractal cross-correlation
between two time series.

\subsubsection{Performance comparison of MFCCA and MF-DXA}

\begin{figure}
\includegraphics[width=0.4\textwidth,height=0.37 \textwidth, clip=true]{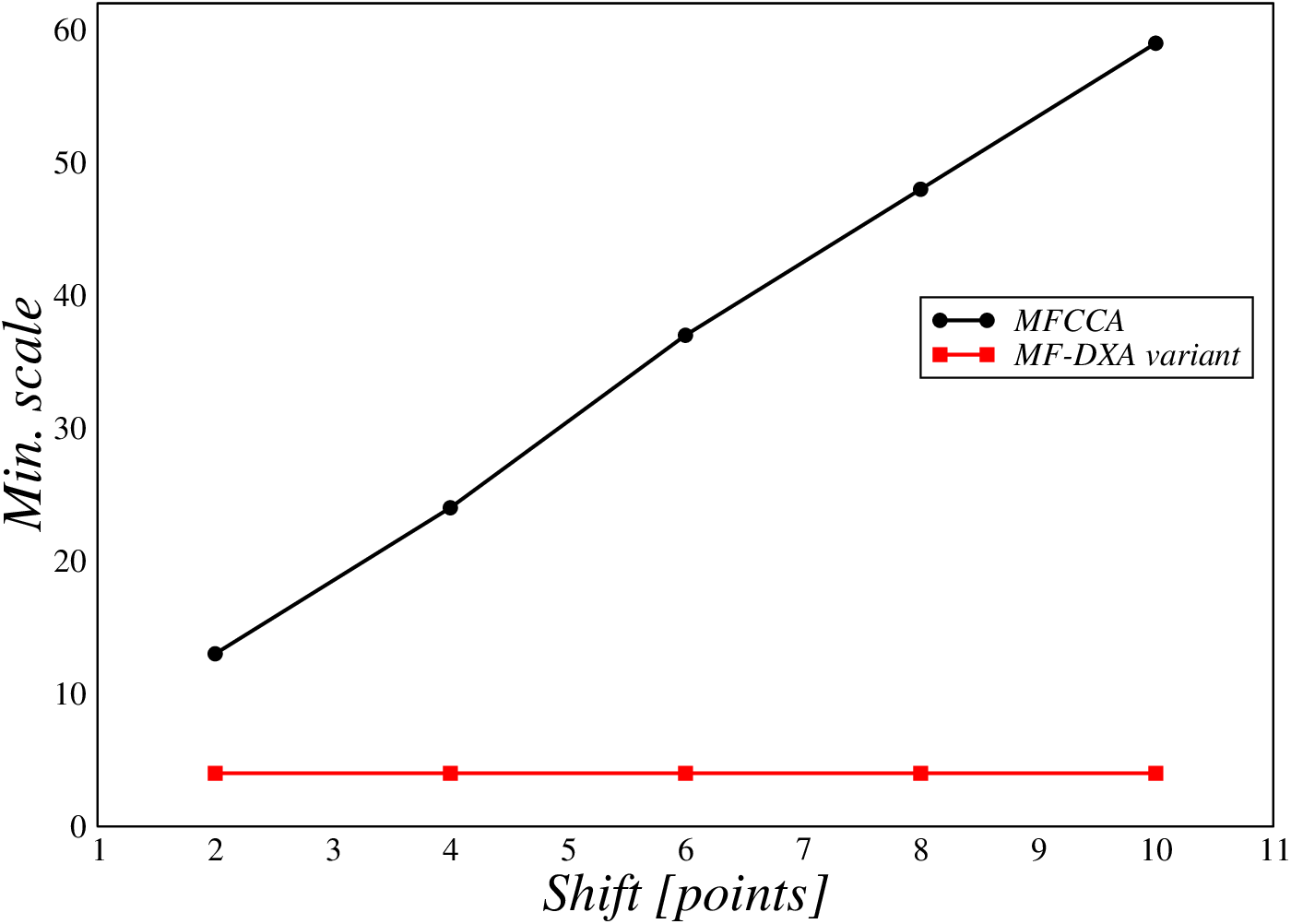}
\caption{(Color online) Lower bound of the scaling regime that can be used in the calculation
of $\lambda_q$ as a function of the shift between two independent realizations
of the MSM time series with $m_0=1.2$. Black circles and red squares refer to
calculations performed by means of MFCCA and a commonly adopted 'modulus'
variant of the MF-DXA procedure, respectively.}
\label{fig8}
\end{figure}

In the next stage of our study of the MSM generated $\sigma_t$ time series, we
analyze output of MFCCA if one time series of a pair is gradually being
shifted in time with respect to the other one. Then the correlations,
especially their fractal character, should undergo an obvious weakening. This
test is aimed at further verifying performance of the algorithm. As an input,
we use a time series with $m_0=1.2$ and the same one, but shifted by a certain
number of points. We notice that the larger is the relative shift between the
time series, the shorter is the scaling range of $F_{xy}$. However, in all
cases, the estimated $\lambda$ is equal to the generalized Hurst exponent
calculated for a single series. This shortening of the range of scaling is not
symmetric from both sides of the scale range, but gradually arrives entirely
from the small scale side. The shift dependence of the lower bound of the
scaling regime that can be used to determine $\lambda_q$ is shown in
Fig.~\ref{fig8}. As expected, lifting of this lower bound is seen to be almost
linear. In the same Figure, we also present the result of an analogous
analysis, but performed by means of the common variant of the MF-DXA procedure
that, in order to resolve the sign problem, makes use of the absolute values
of the fluctuation functions~\cite{jiang11,he11,li12,wang12}. In this case,
the procedure is seen not to be sensitive to this type of surrogate and, thus,
evidently generates spurious cross-correlations.

\begin{figure}
\includegraphics[width=0.51 \textwidth,height=0.26 \textwidth,
clip=true]{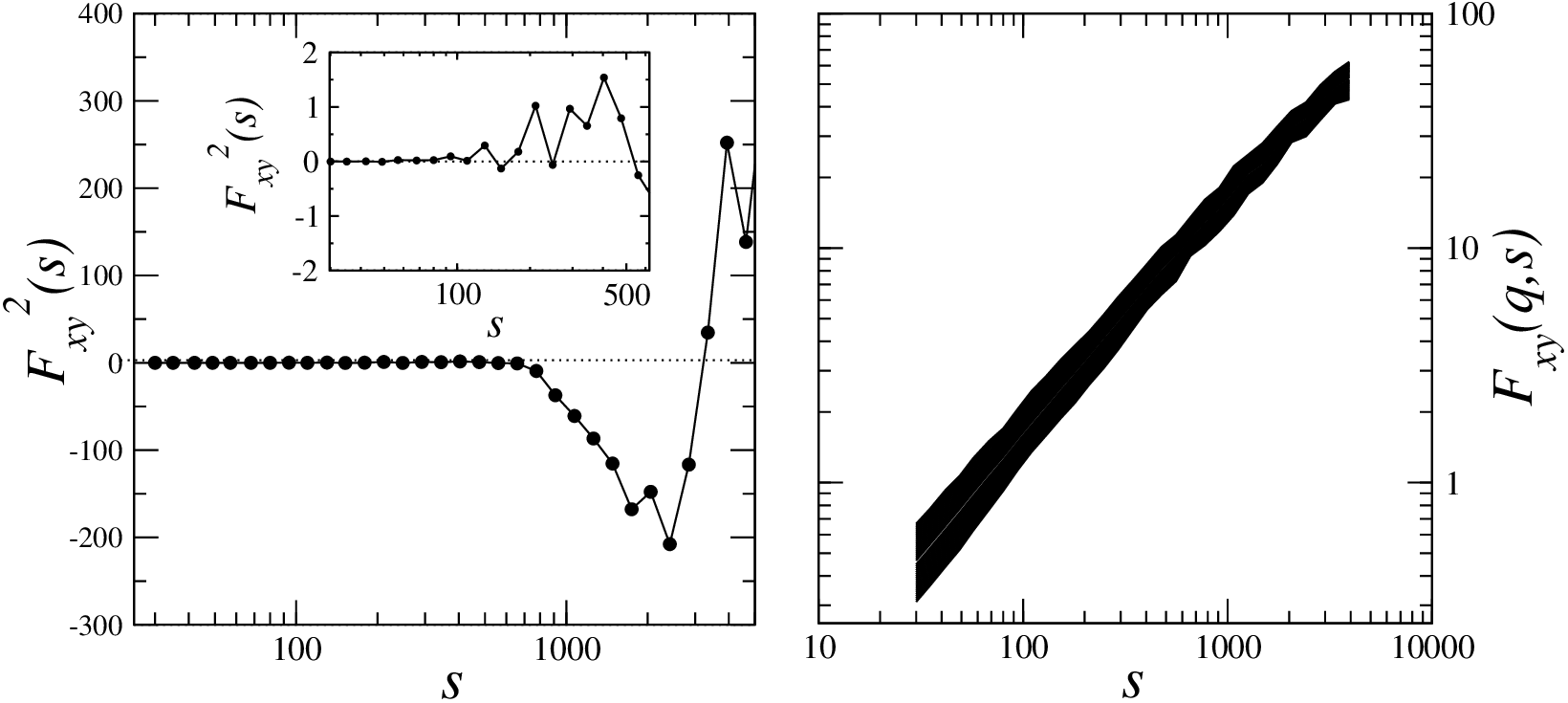}
\caption{(Left) 2nd-order covariance function $F_{xy}^2(s)$ obtained by means
of MFCCA for two independently generated MSM time series with $m_0=1.2$.
(Inset) The same function for a shorter range of scales $s$. (Right) A family
of the $F_{xy}(q,s)$ functions  calculated  by means of the 'modulus' variant
of MF-DXA. The lowest and the highest line refers to $q=-4$ and $q=4$,
respectively.}
\label{fig9}
\end{figure}

In order to elaborate more in detail on this last issue, we generate an
example of a pair of the MSM time series with $m_0=1.2$ drawn independently,
i.e., with no taking care about preserving the hierarchical structure of the
multipliers. Even though individually both such series are multifractal with
the same multifractality characteristics, there is no reason to expect them to
be multifractally cross-correlated. Indeed, in the present case the
corresponding $q$th order covariance functions determined through the
Eq.~(\ref{Fq}) do not scale and for small and moderate scales they even assume
the negative values by fluctuating around zero. An example of $F^2_{xy}(s)$
demonstrating this behavior is shown in the left panel of Fig.~\ref{fig9}. In
fact, this dependence closely resembles the DCCA result (Fig.~1b in
ref.~\cite{podobnik08}) obtained in an analogous situation of the two
uncorrelated series. This correspondence thus provides an additional argument
that it is MFCCA proposed here that constitutes a natural and correct
multifractal generalization of DCCA. Application of a previously
postulated~\cite{zhou08} extension of DCCA in the present example would lead
to complex-valued $q$th-order covariances. As already mentioned in
Introduction, a commonly adopted resolution to this difficulty is based on
taking modulus of the cross-covariance before computing its $q$th order. The
result of such a procedure applied to our example of two independently
generated MSM time series with  $m_0=1.2$ is shown in the right panel of
Fig.~\ref{fig9} and it clearly indicates a convincing multifractal scaling.
This, of course, is however a false signal as these series are not expected to
be multifractally correlated.

\subsubsection{Signed version of the MSM model}

\begin{figure}
\includegraphics[width=0.46 \textwidth,height=0.34 \textwidth,
clip=true]{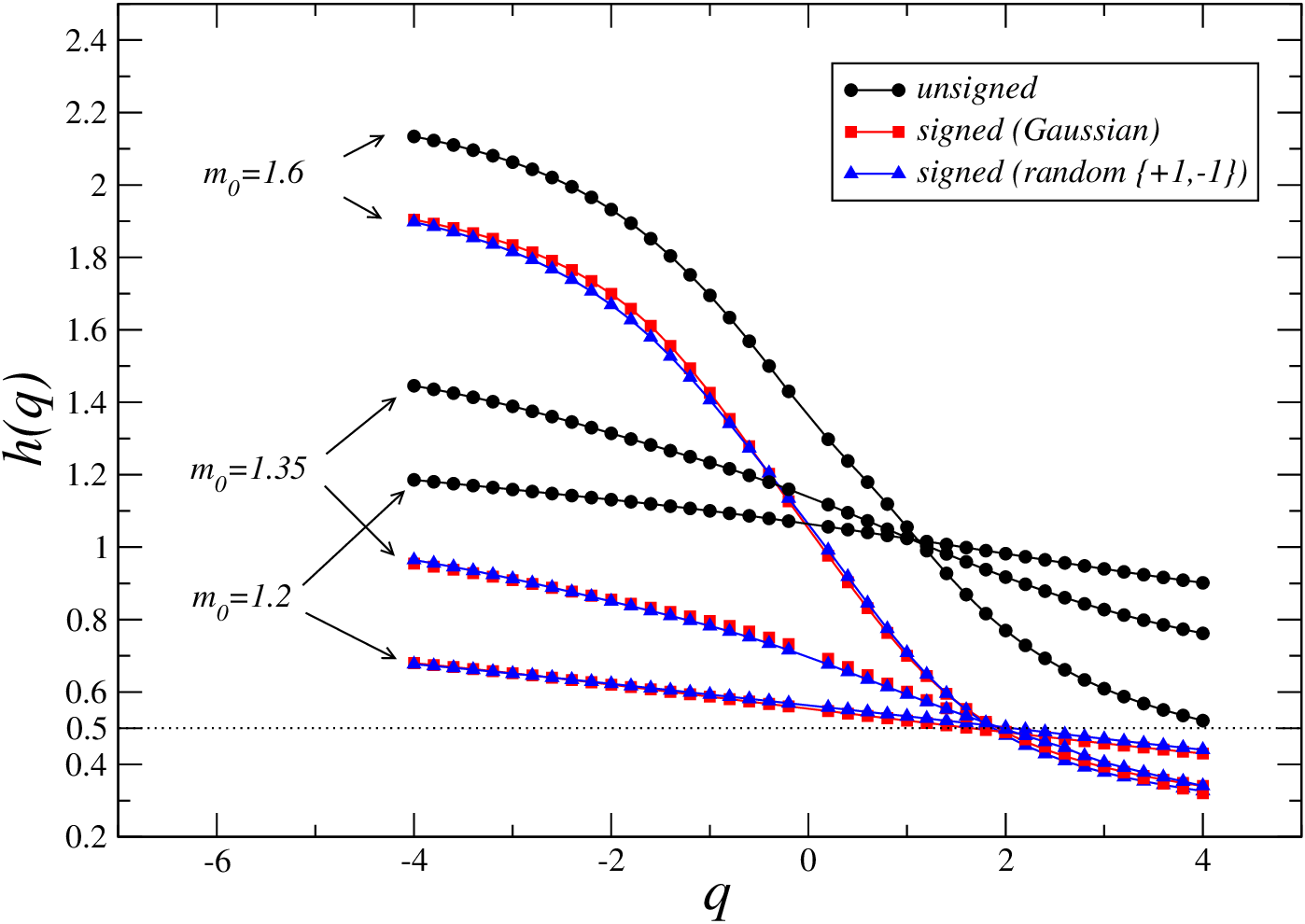}
\caption{(Color online) Generalized Hurst exponent $h(q)$ calculated for individual MSM time
series. The circles refer to the original unsigned series while squares and
triangles refer to the series signed by the Gaussian random variable and by
pure random sign, respectively.}
\label{fig10}
\end{figure}

\begin{figure}
\includegraphics[width=0.5 \textwidth,height=0.33 \textwidth,
clip=true]{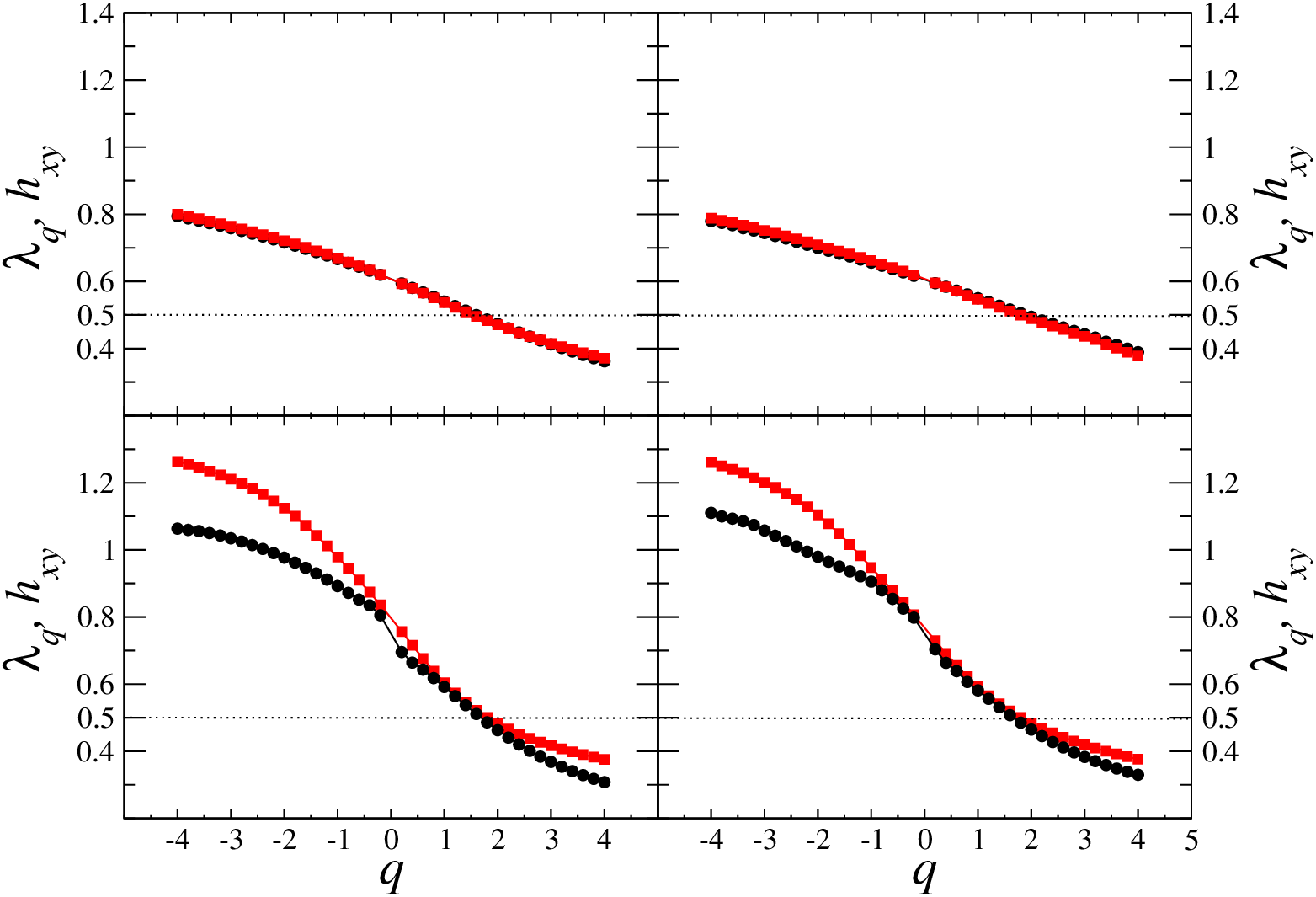}
\caption{(Color online) $q$th-order fluctuation function scaling exponent $\lambda_q$ (black
circles) and the average generalized Hurst exponent $h_{xy}(q)$ (red squares)
calculated for two pairs of signed MSM time series corresponding to
$m_{0}^{(1)}=1.2$, $m_{0}^{(2)}=1.35$ (top) and $m_{0}^{(1)}=1.2$,
$m_{0}^{(2)}=1.6$ (bottom). (Left) MSM time series signed by a Gaussian random
variable. (Right) MSM time series signed by a pure random sign.}
\label{fig11}
\end{figure}

The time series of $\sigma_t$ considered above represent unsigned fluctuations
(volatility in financial terms) and therefore their individual Hurst exponents
are significantly larger than 0.5. By incorporating the Gaussian random
variable $u_t$ drawn from $N(0,1)$ through the Eq.~(\ref{msm}), one obtains
the signed time series $r_t$ with the Hurst exponents close to 0.5 (as for the
financial returns, for instance). Very similar effect one obtains when
multiplying the original unsigned fluctuations simply by randomly drawn either
$+1$ or $-1$. The influence of such procedures on the generalized Hurst
exponents $h(q)$ is shown in Fig.~\ref{fig10} for the same pairs of the MSM
time series as before, i.e., with $m_0=1.2, m_0=1.35$ and $m_0=1.2, m_0=1.6$.
Circles indicate $h(q)$ for the original unsigned series while squares and
triangles indicate the series signed by the Gaussian random variable and by
the pure random sign, respectively. Introducing sign clearly shifts the lines
down relative to the unsigned case, such that the usual Hurst exponent
$H=h(2)$ assumes value of $0.5$ for all the signed series. The $q$-dependence
of $h(q)$, naturally stronger for larger $m_0$, remains however essentially
preserved after introducing the sign, which reflects the fact that such an
operation influences primarily the linear temporal correlations in the series
leaving the nonlinear ones, related to the volatility
clustering~\cite{drozdz09}, preserved. As far as multifractal
cross-correlations between such series are concerned, more care is needed.
Drawing the term $u_t$ in Eq.~(\ref{msm}) independently for the two series
destroys their original (unsigned) cross-correlations and the corresponding
$q$th-order covariances calculated through the Eq.~(\ref{Fq}) develop similar
fluctuations as those in the left panel of Fig.~\ref{fig9}. One most
straightforward way to preserve multifractal cross-correlations is to use the
same $u_t$ for the two series under consideration. Examples of the so-prepared
pairs of series, for the same combination of the parameters $m_0$ as before
for the unsigned series, i.e., $m_0=1.2$ versus $m_0=1.35$ and $m_0=1.2$
versus $m_0=1.6$, are analyzed in Fig.~\ref{fig11} in terms of $\lambda_q$ and
$h_{xy}(q)$. For the first of these pairs, irrespective of the sign adding
variant, the multifractal cross-correlations are seen to remain essentially on
the same level of strength as those for the corresponding unsigned signals
shown in the upper panel of Fig~\ref{fig2}. The departures between $\lambda_q$
and $h_{xy}(q)$ for the other pair ($m_0=1.2$ and $m_0=1.6$) of the signed
series can be seen to be somewhat larger relative to their unsigned
counterparts, which signals slight weakening of their multifractal
cross-correlations. This in fact is consistent with the generalized Hurst
exponents $h(q)$ seen in Fig.~\ref{fig10}. When sign is applied to the series,
the distance between the corresponding $h(q)$ routes increases, especially on
the negative $q$ side, as compared to the unsigned case.

\subsection{Examples of stock market data \label{stocks}}

The financial fluctuations can be considered a physical process which
constitutes one of the most complex generalizations of the conventional
Brownian motion carrying at the same time convincing traces of nontrivial
fractality~\cite{mandelbrot97,bogachev07,ludescher11b}. They therefore offer a
very demanding territory to test the related concepts and algorithms. For this
reason, as final examples of utility of the MFCCA method, we present an
analysis of empirical data coming from the German stock exchange. Furthermore,
since multifractal analysis of financial data is one of the most informative
methods of investigating such complex
systems~\cite{oswiecimka05,drozdz10,zhou09,kwapien05,lopez03}, we believe that
MFCCA will be very useful in this field as well. We consider logarithmic price
increments $g(i)$ and linear time increments $\Delta t(i)$ representing
dynamics of a sample German stocks - E.ON (ticker: EOA) and Deutsche Bank
(ticker: DBK) (from the same database as used before~\cite{oswiecimka05})
being part of the DAX30 index. These quantities are obtained according to the
formulas:
\begin{equation}
g(i)=\ln(p(i+1))-\ln(p(i)), \ \
\end{equation}
\begin{equation}
\Delta t(i)=t(i+1)-t(i),
\end{equation}
where $p(i), i=1,...,N$ is a time series of price quotes taken in discrete
transaction time $t(i)$. As it has been shown previously~\cite{oswiecimka05},
both $g(i)$ and $\Delta t(i)$ are processes with self-similar structure and
could be analyzed by the multifractal methods. Quantifying the character of
cross-correlations just between these two characteristics of the financial
dynamics is also of particular importance for forecasting volatility within
models such as the Multifractal Model of Asset
Returns~\cite{mandelbrot96,mandelbrot97,lux03,oswiecimka06b,eisler04}. Our
analysis is performed on time series comprising the period between Nov.~28,
1997 and Dec.~31, 1999. The time series consists of $T=294,862$ and
$T=497,513$ points for EOA and DBK, respectively. Therefore, the time series
are long enough to bring statistically significant results.

\begin{figure}
\includegraphics[width=0.45\textwidth,height=0.35 \textwidth, clip=true]{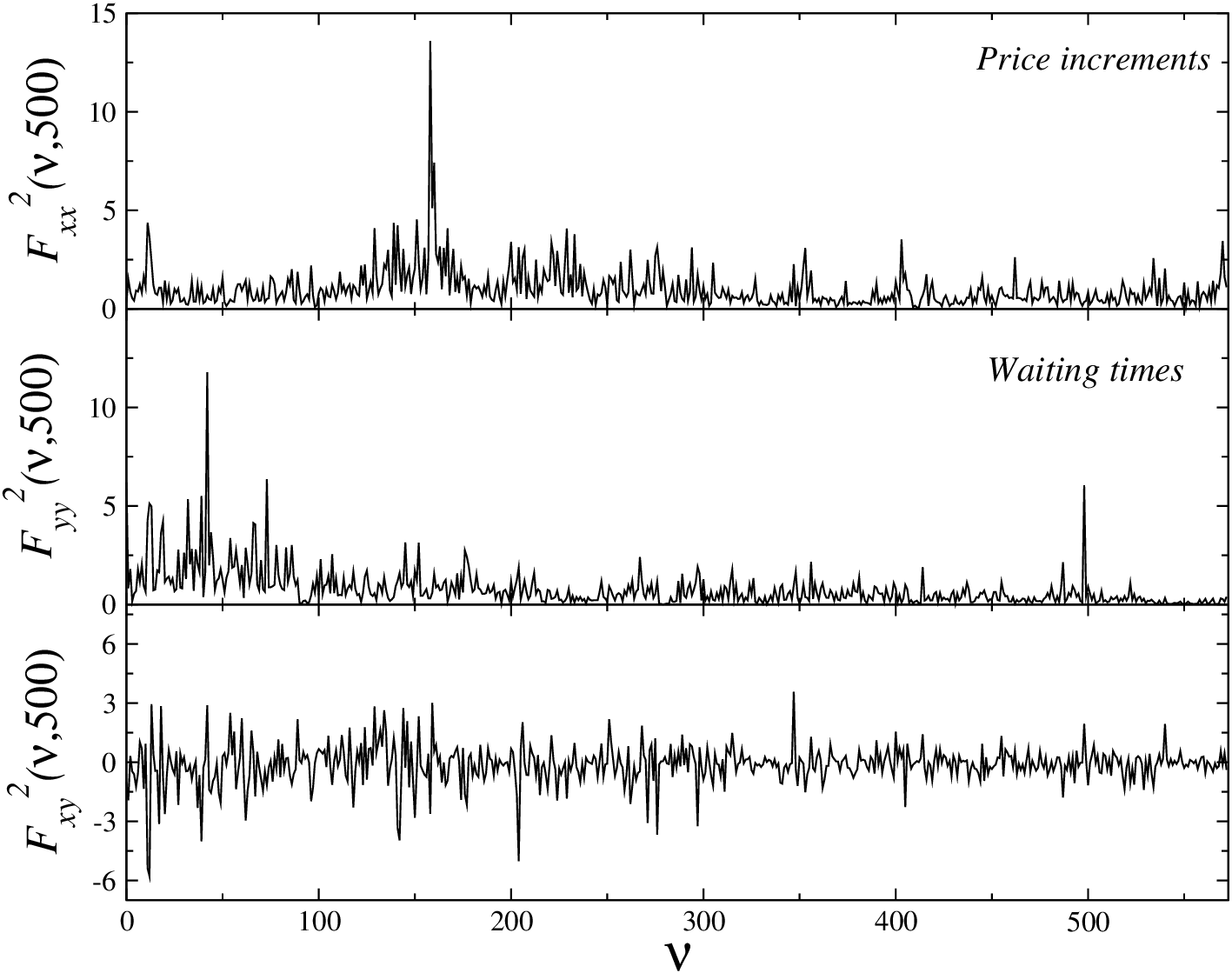}
\caption{Top and middle: detrended variance functions $F_{xx}^2(\nu,500)$ and
$F_{yy}^2(\nu,500)$ calculated for time series of price increments and waiting
times of E.ON stock (ticker: EOA), respectively. Bottom: detrended
cross-covariance function $F_{xy}^2(\nu,500)$ calculated for the same data.
Calculations were carried out for segments of length of 500 points.}
\label{fig12}
\end{figure}

In Fig.~\ref{fig12}, we show one of the first steps of our algorithm, i.e.,
the detrended cross-covariance function $F^2_{xy}(\nu,500)$ (for the scale
$s=500$) as a function of the box number $\nu$ (Eq.~(\ref{Fxy2})). For
comparison, in the same Fig.~\ref{fig12}, we depicted the detrended variance
function $F_{xx}^2(\nu,500)$ and $F_{yy}^2(\nu,500)$ (obtained from MFDFA)
calculated for individual time series of price increments and waiting times,
respectively. It is easy to notice that the detrended variance calculated for
individual time series takes only positive values, whereas the detrended
cross-covariance function $F_{xy}^2(\nu,s)$ takes both negative and positive
values. This constitutes the already-mentioned principal problem in
straightforward calculation of the $q$th-order cross-covariance function
$F_{xy}(q,s)$ for odd $q$s that results in complex values of this function
(see Eq.~(\ref{Fxy})). It is worth stressing that this difficulty does not
affect the fractal analysis of individual time series (MFDFA), because then
the detrended variance function $F_{xx}^2(\nu,s)$ may only be positive. It
follows that proper handling of the sign of $F_{xy}^2(\nu,s)$ is of crucial
importance for a consistent extension of DCCA to treat the multifractally
correlated signals. At present, a solution of this problem is offered only by
the MFCCA algorithm proposed in Sec.~\ref{Method}.

\begin{figure}
\includegraphics[width=0.5\textwidth,height=0.4 \textwidth,
clip=true]{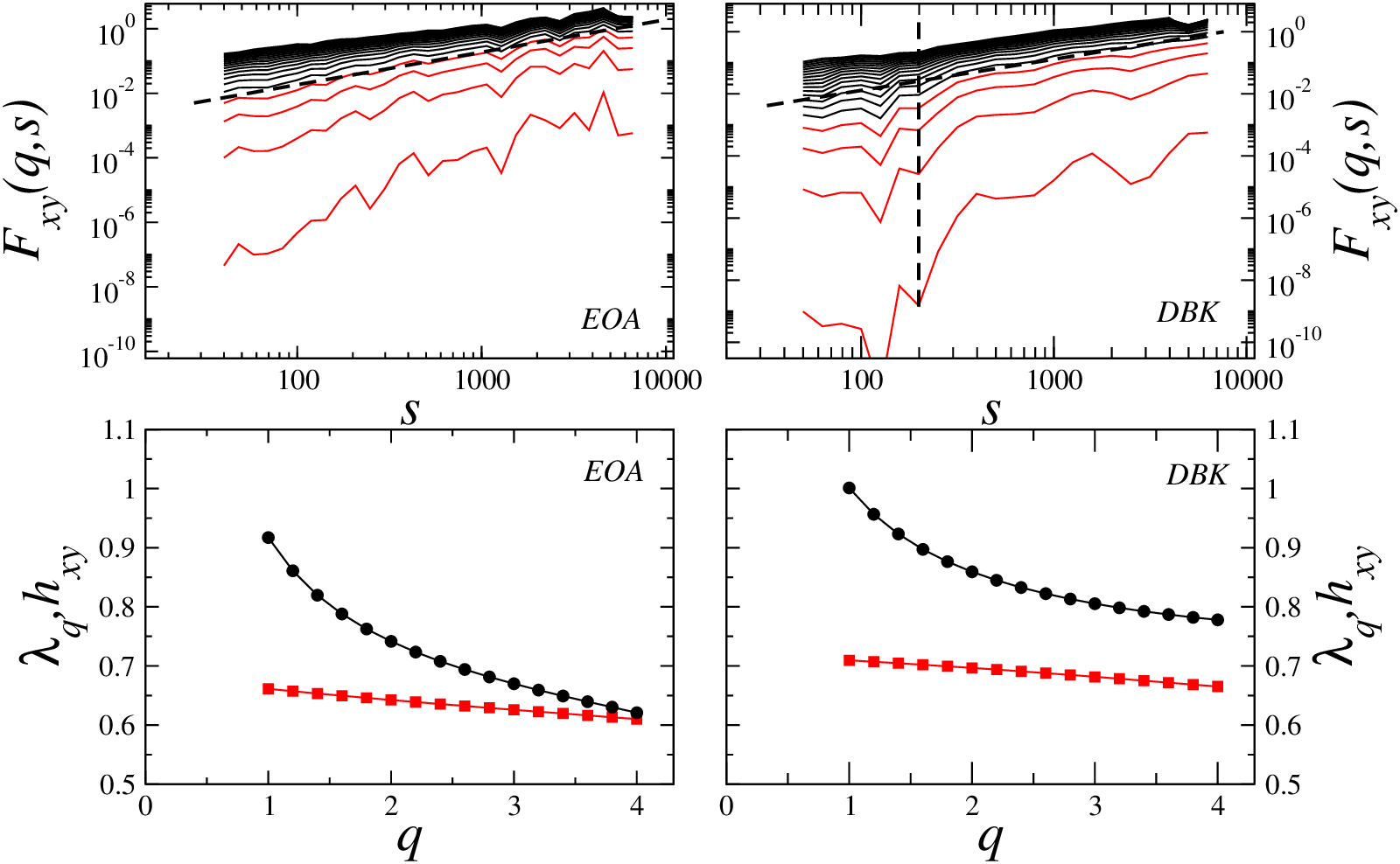}
\caption{(Color online) (Top) Family of the $F_{xy}(q,s)$ functions calculated for time
series of price increments and waiting times of E.ON (EOA) and Deutsche Bank
(DBK). The lowest and the highest line on each panel refers to $q=0.2$ and
$q=4$, respectively. The dashed lines indicate scaling bounds both in $q$ and
$s$. (Bottom) Multifractal cross-correlation exponent $\lambda_q$ (black
circles) and the average generalized Hurst exponents $h_{xy}(q)$ (red squares)
for the same data as above. The exponents $\lambda_q$ are estimated only for
$1 \le q \le 4$.}
\label{fig13}
\end{figure}

In order to characterize the cross-correlations in the present case, the
function $F_{xy}(q,s)$ is calculated. As far as the multifractal scaling is
concerned, the situation is significantly more subtle than in the previous
model cases. It turns out that the scaling property of $F_{xy}(q,s)$ applies
only selectively. First of all, for the negative $q$s, $F_{xy}^q(s)$
fluctuates around zero and Eq.~(\ref{Fxy}) is not satisfied. For positive
values of $q$, the function $F_{xy}^q(s)$ assumes positive values but
clear scaling of $F_{xy}(q,s)$ begins with $q=1$ upwards. For $q < 1$, these functions develop
increasing fluctuations when $q$ moves towards zero. This effect is especially
strong for DBK. Furthermore, the lower limit of scales where $F_{xy}(q,s)$
develops the convincing power-law behavior varies and it takes place at the
higher values of $s$ for DBK than for EOA, which signals a weaker form of
multifractal cross-correlation in the former case. The corresponding
characteristics are shown in the upper panels of Fig.~\ref{fig13} with the
scaling bounds both in $q$ and in $s$ indicated by the dashed lines.

\begin{figure}
\includegraphics[width=0.45\textwidth,height=0.35 \textwidth,
clip=true]{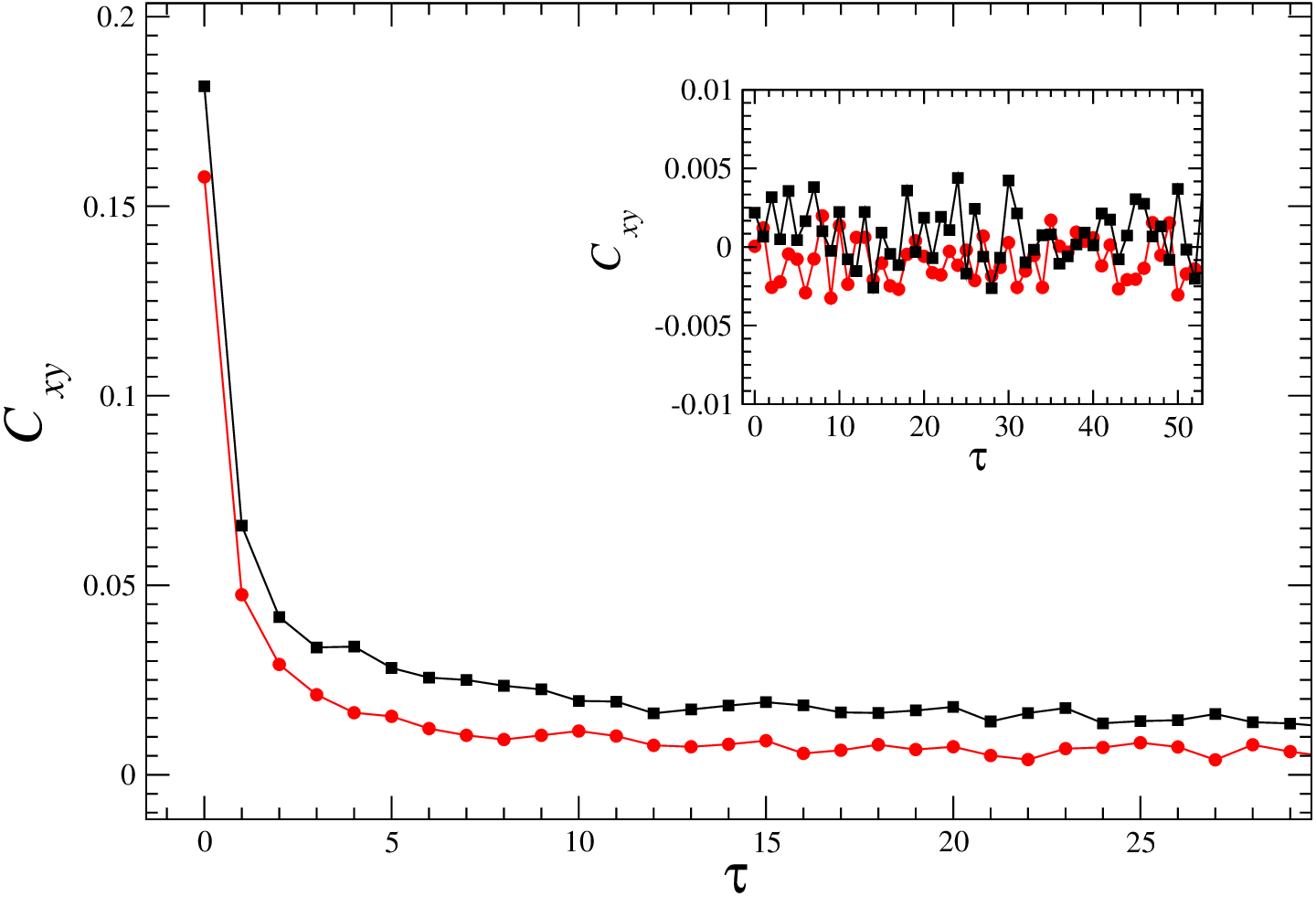}
\caption{(Color online) Cross-correlation function $C_{xy}(\tau)=|x_{i+\tau}||y_i|$
corresponding to modulus of price increments and transaction times of EOA
(black squares) and DBK (red circles). Inset: the same function but calculated
for randomly shuffled data.}
\label{fig14}
\end{figure}

The calculated $\lambda_q$ and $h_{xy}(q)$ are shown in the bottom panels of
Fig.~\ref{fig13}. It is clearly visible that, for EOA, both functions converge
to each other for large values of $q$, while $\lambda_q$ is significantly
larger than $h_{xy}(q)$ for smaller values of $q$. These results imply that
the scaling properties of $F_{xy}(q,s)$ strongly depend on the considered time
span and they cannot be fully quantified by a unique exponent $\lambda$.
Moreover, based on our results for the MSM model, we can infer that the
analyzed processes are ruled by the similar fractal dynamics only in periods
with relatively large $F_{xy}^2(\nu,s)$ (associated with large $q$). For
smaller $q$'s, the difference between $\lambda_q$ and $h_{xy}(q)$ is more
evident, which suggests that dynamics of these processes is significantly
different, but still cross-correlative. It is worth to mention that large
values of $F_{xy}^2(\nu,s)$ can be a consequence of cross-correlation both in
the signs and the amplitudes of the signals. However, the waiting times are
unsigned and the price increments are signed, but the sign is uncorrelated.
This means that, in our case, the amplitude of $F_{xy}^2(\nu,s)$ is only a
result of the cross-correlation of the observed amplitude. The strong
cross-correlation of volatility (modulus of time series) is confirmed by
Fig.\ref{fig14}, where the cross-correlation function for the waiting times
and absolute values of the price increments is depicted. Therefore, we
conclude that large fluctuations are much more strongly cross-correlated than
the smaller ones. Complexity of the multifractal cross-correlation is
expressed by the range of $\lambda_q$ that is approximately $0.32$ in this case.

As may be anticipated already from the structure of $F_{xy}(q,s)$ for DBK, the
behavior of $\lambda_q$ is slightly different and the difference between
$h_{xy}(q)$ and $\lambda_q$ is substantial both for small and for large values
of $q$ (Fig.~\ref{fig13}, right panel). Also $\Delta \lambda_q$ for DBK is
smaller than in the case of EOA and takes a value of $0.22$. This suggest that
although structure of the cross-correlation between the inter-transaction
times and the price increments for DBK is multifractal, its heterogeneity is
poorer than in the case of EOA. Moreover, similarity between the fractal
dynamics of large fluctuations is not so evident than in the former case.
These results are also confirmed by Fig.~\ref{fig14}, where a difference
between the strength of volatility cross-correlations for both considered
stocks is easily visible.

The results presented here indicate that the multifractal cross-correlation
characterizes only relatively large fluctuations of the signals under study.
Smaller fluctuations that are filtered out by $q<1$, from the perspective of
multifractal cross-correlation, may be considered mutually independent.

In connection with the present example we also wish to mention - but without
showing the results explicitly in order not to confuse the reader - that
taking absolute values of the fluctuation functions to get rid of the sign
problem (as recently often done in literature~\cite{he11,li12,wang12}), in the
present financial data case, would result in a convincing but apparent
multifractal scaling for all values of $q$, similar to one that we have
already seen for the MSM model in Fig.~\ref{fig9}. Also, the so determined
$\lambda_q$ equals $h_{xy}(q)$ as in the MSM model. This way one, however,
does not extract genuine correlations, but only measures the averaged
multifractal properties of individual time series.

\begin{figure}
\includegraphics[width=0.5\textwidth,height=0.4 \textwidth, clip=true]{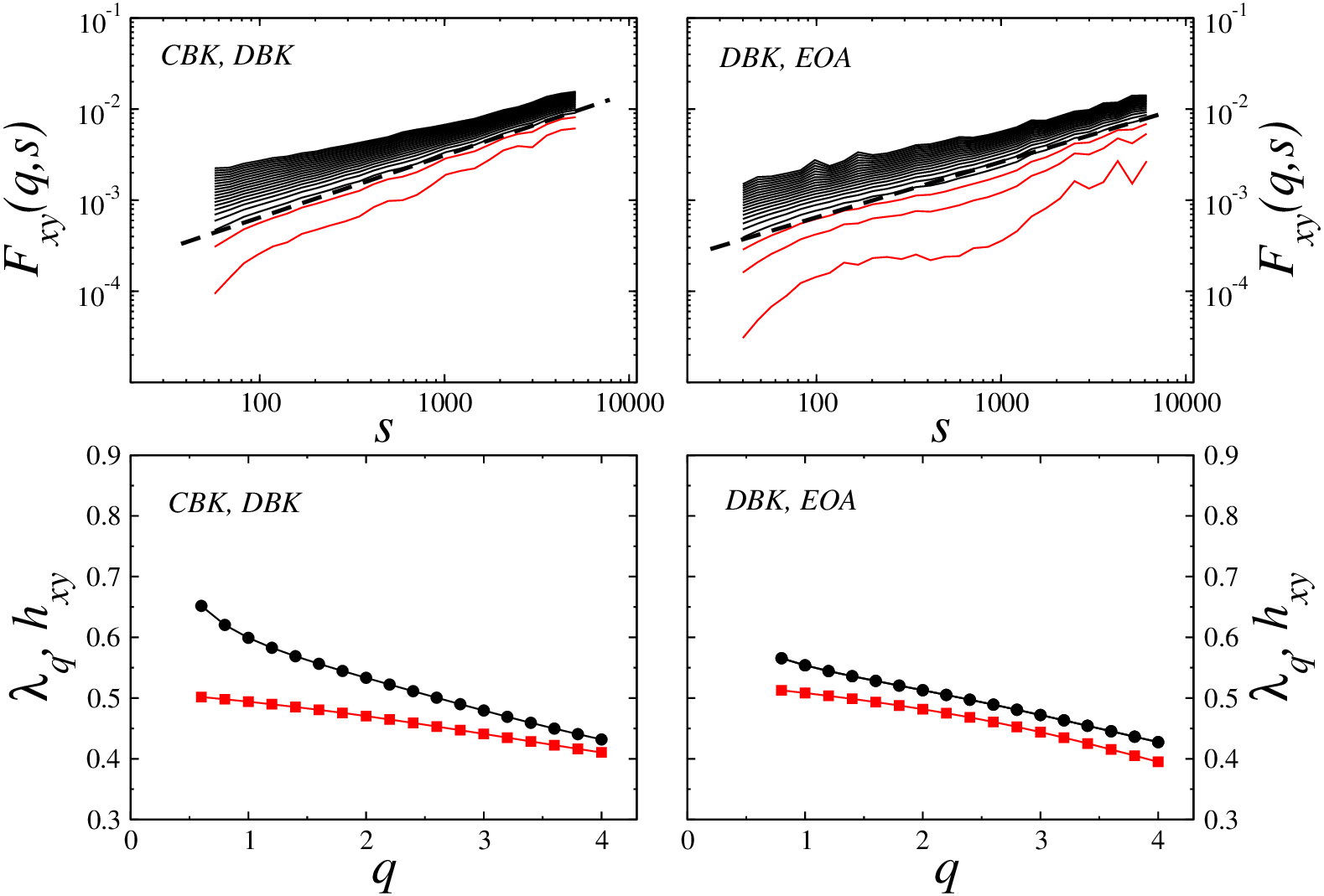}
\caption{(Color online) (Top) Family of the $q$th-order cross-covariance functions
$F_{xy}(q,s)$ calculated for 1 min returns of two pairs of stocks
corresponding to Commerzbank (CBK) and Deutsche Bank (DBK), and Deutsche Bank
and E.ON (EOA). The lowest and the highest line on each panel refers to
$q=0.2$ and $q=4$, respectively. The dashed lines indicate scaling bounds both
in $q$ and $s$. (Bottom) Multifractal cross-correlation exponent $\lambda_q$
(black circles) and the average generalized Hurst exponents $h_{xy}(q)$ (red
squares) for the same data. Calculation of $\lambda_q$ was restricted to $0.6
\leq q \le 4$.}
\label{fig15}
\end{figure}

Another type of correlations that are of theoretical as well as of practical
interest are the correlations among stock returns~\cite{kwapien12,mantegna00}.
These are typically quantified in terms of the Pearson correlation
coefficients or, more generally, in terms of the correlation matrix. This way
of quantifying correlations is, however, restricted to their linear component
only. The present formalism of studying the multifractal cross-correlations
allows one to reveal some of their potential nonlinear components. As an
example, we therefore use the same two stocks as above (EOA and DBK) and, in
addition, Commerzbank (CBK) from the same, German stock exchange over the same
period, and perform a similar analysis as above for two pairs  of time series
(CBK-DBK and DBK-EON) representing the corresponding 1 min returns. Over the
period considered, this yields 267,241 data points. The results in the same
representation as before are presented in Fig.~\ref{fig15}.

For $q < 0$, the $F_{xy}(q,s)$ are not drawn since the corresponding
$F_{xy}^q(s)$ functions fluctuate around zero. As we go to the positive $q$
values, however, they start developing a convincing scaling already for
$q=0.6$ (as indicated by the dashed lines) for both pairs and for all the
scales considered. This scaling is clearly multifractal and the resulting
$\lambda_q$ and $h_{xy}(q)$, shown in the lower panels of Fig.~\ref{fig15},
are somewhat closer to each other than the ones previously considered for
correlations between the price increments and the inter-transaction times.
Slight differences in relation between the present two pairs of time series
are also visible, however. For DBK-EON, the departures between $\lambda_q$ and
$h_{xy}(q)$ are largely independent on $q$ in the region where scaling
applies, while for CBK-DBK it starts from larger values for the smallest
$q$-values (0.6), but it converges to even smaller values with an increasing
$q$. This can be interpreted as an indication that multifractal character of
cross-correlations resembles more each other for CBK and DBK on the level of
large fluctuations and weakens for the smaller ones, while, within the pair
DBK-EON, they are of similar strength in the comparable range of fluctuation
size. Of course, in both cases this kind of cross-correlations disappears on
the level of small fluctuations that are filtered out by the negative values
of $q$, and this seems quite a natural effect in the financial context.

\begin{figure}
\includegraphics[width=0.45\textwidth,height=0.45 \textwidth,
clip=true]{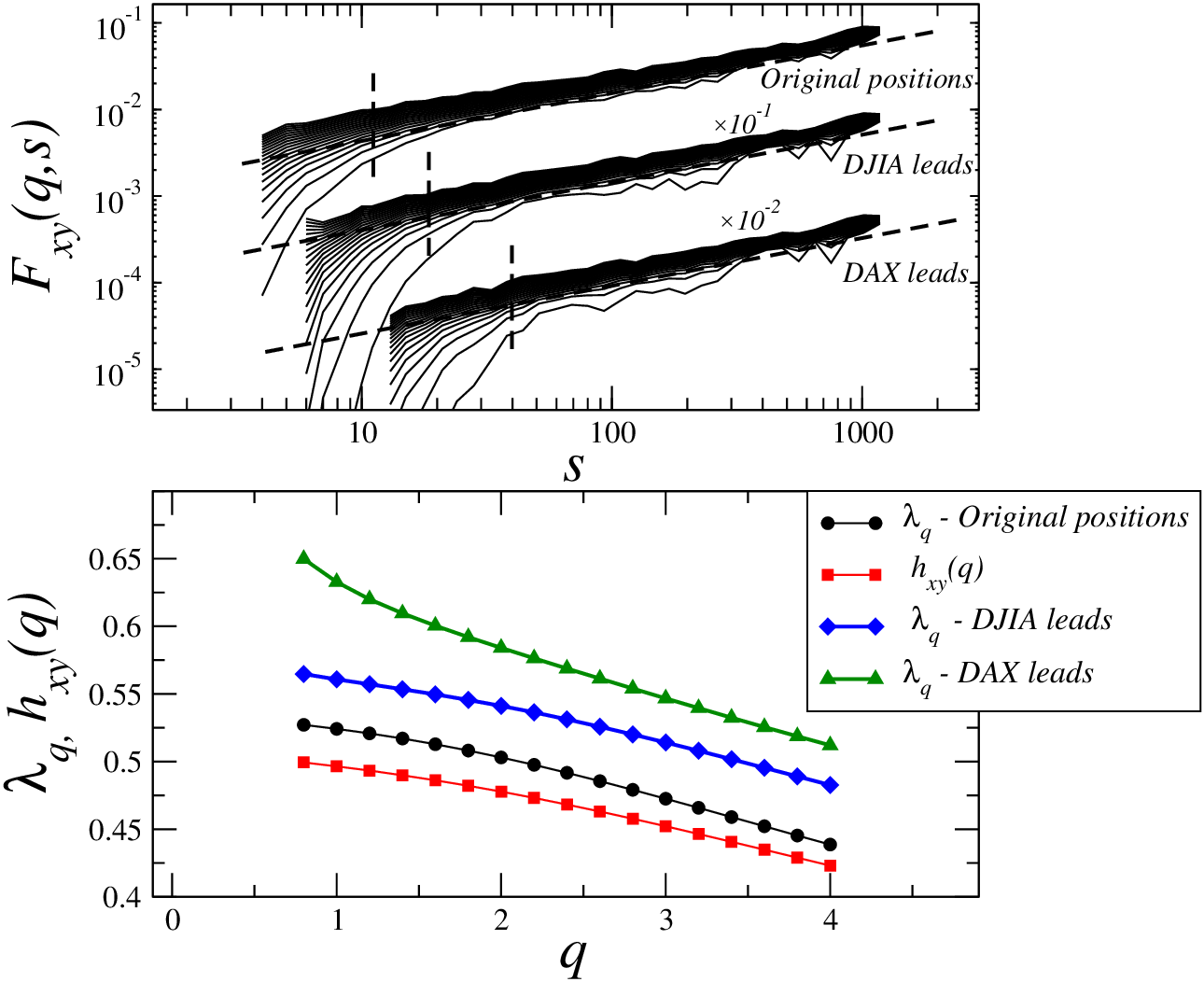}
\caption{(Color online) (Top) Three families of the $q$th-order cross-covariance functions
$F_{xy}(q,s)$ calculated for different synchronization levels of time series
representing daily returns of two market indices, DJIA and DAX: (1) the
synchronous (original) index positions, (2) DAX retarded by one day with
respect to DJIA (DJIA leads), and (3) {\it vice versa} (DAX leads). For
clarity, the functions $F_{xy}(q,s)$ for 'DJIA leads' and 'DAX leads' are
vertically shifted. The lowest and the highest line for each considered case
refers to $q=0.2$ and $q=4$, respectively. The dashed lines indicate scaling
bounds both in $q$ and $s$. (Bottom) Multifractal cross-correlation exponent
$\lambda_q$ and the average generalized Hurst exponents $h_{xy}(q)$ calculated
for the corresponding time series ($0.8 \leq q \leq 4$).}
\label{fig16}
\end{figure}

As a final example indicating possible applications of the MFCCA method
introduced in this paper, we study the cross-correlations between the two
world leading stock market indices, the Dow Jones Industrial Average (DJIA)
and the Deutscher Aktienindex (DAX), based on their daily returns. The period
considered for both these indices begins on January 12, 1990 and ends on
October 12, 2013. This results in time series of length of 5881 data points.
Due to different time zones which the two indices are traded in and in order
to test potential applicability of the present algorithm in detecting possible
time-lags or asymmetry effects in correlations, we study three possible
variants of positioning the time series relative to each other. The first
variant is most natural, i.e., data points in the two time series meet each
other at the same date they are recorded. The other two variants are such that
the time series are shifted by one day relative to each other, either DAX is
advanced by one day or DJIA is. The corresponding $F_{xy}(q,s)$ functions are
displayed in the upper panel of Fig.~\ref{fig16}. Unlike the high-frequency
recordings discussed above, the significantly shorter time series in the
present case restrict us to cover a smaller scale range. Nevertheless, evident
multifractal scaling can still be identified in this case as indicated by the
dashed lines in Fig.~\ref{fig16}, provided the range of $q$ is restricted as
well. Similarly to the situation with the high frequency cross-correlations
within the German stocks, here $F_{xy}^q(s)$ also fluctuates around zero for
negative $q$s, and therefore the corresponding functions $F_{xy}(q,s)$ are not
shown. Interestingly, the lower bound in $s$ where scaling starts visibly
lifts up as we move from the same date, through the situation described as
'DJIA leads', and becomes the shortest in the situation 'DAX leads'.
Accordingly, the departures between $h_{xy}(q)$, which, of course, remains
invariant with respect to such relative shifts of the time series, and
$\lambda_{q}$ increase as we go through the above three relative locations of
DJIA versus DAX (lower panel of Fig.~\ref{fig16}). The strongest DJIA-DAX
multifractal cross-correlation is detected when the series are originally
arranged relative to each other. Their relative 1-day shifts reveal an effect
of asymmetry, however. The situation 'DJIA leads' preserves significantly more
of such cross-correlations than the opposite "DAX leads' one. This result can
be interpreted as an indication that the DJIA close has more influence on the
DAX close next day than the DAX close has on the DJIA close next day. In fact,
as verified additionaly, splitting the time series considered here into two
halves shows that this effect is more evident in 1990's than more recently.
Such an asymmetry in information transfer between these two stock markets is
understandable in economic terms, and in fact it is also consistent with the
previous study~\cite{drozdz01} based on the correlation matrix formalism.
Finally, we wish to mention that more distant relative shifts of the two
present time series quickly deteriorate the multifractal cross-correlation,
while at the same time the modulus-based MF-DXA approach leaves them unchanged.

\section{Summary and conclusions}

We proposed an algorithm, which we called Multifractal Cross-Correlation
Analysis, that allows for quantitative description of multiscale
cross-correlations between two time series and that is free of limitations the
other existing algorithms, like MF-DXA, suffer from. The key point that
distinguishes MFCCA from other related methods is construction of the
$q$th-order cross-covariance function $F_{xy}^q(s)$ in Eq.(\ref{Fq}), which
preserves the sign of the cross-covariance fluctuation function
$F_{xy}^2(\nu,s)$ after its modulus has been raised to a power of $q/2$. This
step has two immediate consequences: (1) it eliminates the risk of appearance
of complex values that might lead to problems with their correct
interpretation, and (2) it prohibits losing information that is stored in the
negative cross-covariance. It follows that, as we showed in Sec.~\ref{Markov}
regarding known model data, the results obtained with MFCCA are more logical
and better coincide with intuition than do the parallel results of MF-DXA.
This was true both for the signed and the unsigned, volatility-like processes.
On this ground we concluded that MFCCA provides us with the most complete
information about fractal cross-correlations possible as compared to the other 
related methods existing so far.
Having realized this, we applied MFCCA to sample real-world data from the
stock markets. We found that both the cross-stock correlations and the lagged
inter-market correlations of returns, as well as the correlations between
price movements and the corresponding transaction time intervals are clearly
multifractal. Moreover, we showed that carriers of these cross-correlations
are predominantly the large fluctuations in both signals, while the smaller
fluctuations contribute rather little. This outcome may suggest that an
important ingredient of financial complexity, which manifests itself here as
multifractality, might be temporal relations between large events.

Apart from the introduction of MFCCA, we also focused our attention on the
relation between the $q$th-order scaling exponent $\lambda_q$ and the averaged
generalized Hurst exponents $h_{xy}(q)$. Both these measures are equally
important if one intends to comprehend fractal structure of the data under
study. This is because their spectra analyzed in parallel for each signal
separately contain information about similarity of their fractal structure.
For example, based on model data, we found that the larger is the difference
between $\lambda_q$ and $h_{xy}(q)$, the more different are the considered
(multi)fractals. We thus strongly recommend investigation of both these
quantities in parallel.

We believe that our approach presented here will allow for a wider application
of multifractal cross-correlation analysis to empirical data in different
areas of science.

\section{Acknowledgements}

The calculations were done at the Academic Computer Centre CYFRONET AGH,
Krak\'ow, Poland (Zeus Supercomputer) using Matlab environment.

\end{document}